\documentclass[fleqn,usenatbib]{mnras}

\usepackage{newtxtext,newtxmath}

\usepackage[T1]{fontenc}
\usepackage{ae,aecompl}

\usepackage[dvipsnames]{xcolor}
\usepackage{graphicx}
\usepackage{amsmath}
\usepackage{amssymb}
\usepackage{placeins}
\usepackage{color}
\usepackage{microtype}
\usepackage{booktabs}
\usepackage{threeparttable}
\usepackage{tabularx}
\usepackage{enumerate}

\title[Rotational phase dependence of magnetar bursts]{The rotational phase dependence of magnetar bursts}

\author[C.Elenbaas et al.]{
C. Elenbaas,$^{1}$\thanks{E-mail:C.P.C.Elenbaas@uva.nl} A.L. Watts,$^{1}$ and D. Huppenkothen$^{2,3,4}$\\
$^{1}$Anton Pannekoek Institute for Astronomy, University of Amsterdam, Science Park 904, 1098 XH Amsterdam, the Netherlands
\\
$^{2}$DIRAC Institute, Department of Astronomy, University of Washington, Box 351580, Seattle, WA 98195-1580, USA
\\
$^{3}$Center for Data Science, New York University, 65 5h Avenue, 7th Floor, New York, NY 10003, USA
\\
$^{4}$Center for Cosmology and Particle Physics, Department of Physics, New York University, 4 Washington Place, New York, NY 10003, USA
\\
}

\pubyear{2017}

\begin{document}
\label{firstpage}
\pagerange{\pageref{firstpage}--\pageref{lastpage}}

\maketitle

\begin{abstract}
The trigger for the short bursts observed in $\gamma$-rays from many magnetar sources remains unknown. One particular open question in this context is the localization of burst emission to a singular active region or a larger area across the neutron star. While several observational studies have attempted to investigate this question by looking at the phase dependence of burst properties, results have been mixed. At the same time, it is not obvious a priori that bursts from a localized active region would actually give rise to a detectable phase-dependence, taking into account issues such as geometry, relativistic effects, and intrinsic burst properties such brightness and duration. In this paper we build a simple theoretical model to investigate the circumstances under which the latter effects could affect detectability of a dependence of burst emission on rotational phase. We find that even for strongly phase-dependent emission, inferred burst properties may not show a rotational phase dependence depending on the geometry of the system and the observer. Furthermore, the observed properties of bursts with durations short as 10-20\% of the spin period can vary strongly depending on the rotational phase at which the burst was emitted.  We also show that detectability of a rotational phase dependence depends strongly on the minimum number of bursts observed, and find that existing burst samples may simply be too small to rule out a phase dependence.
\end{abstract}


\section{Introduction}		

Magnetars, the most highly magnetized neutron stars (with dipole fields $\gtrsim 10^{13}$ G), are isolated stars powered primarily by magnetic field decay \citep{Thompson1993,Thompson1995,Kouveliotou1998,Kouveliotou1999}.  One of their key characteristics is the sporadic emission of soft $\gamma$-ray bursts \citep[for reviews of this specific aspect see][]{Woods2006,Turolla2015,Kaspi2017}.  Durations and fluences vary, but most of these bursts are short, lasting $\sim 0.01 - 1$ s, less than a typical magnetar spin period $P\sim 6$ s.  The slow decay of the strong magnetic field is assumed to build up stresses in the system: stress release must involve rapid reconfiguration of the external magnetic field, particle acceleration, and $\gamma$-ray emission. However what triggers the occurrence of individual bursts, the way in which a burst progresses, and the associated emission processes, remain very poorly understood \citep{Turolla2015}. The failure point could be internal, within the crust of the star, or in the external magnetosphere itself. 

One question is whether the bursts are triggered within a specific active region, fixed in the rotating frame of the star. Some crust zones, for example, are expected to be particularly prone to magnetically-induced faulting and yielding \citep[see for example][]{Lander2015,Gourgouliatos2015,Thompson2017}. Certain regions of the magnetosphere could also be more active than others, in which case there may also be a preferred height above the neutron star surface. Being able to identify whether this is the case would certainly help in efforts to determine the burst mechanism. 
 
One way to determine this, suggested by \citet{Lyutikov2002}, is to look at whether there is any rotational phase dependence to the bursts. A number of observational studies have attempted to investigate this, using various different measures such as the phase-dependence of the time at which the burst peak is recorded, or the phase-distribution of all of the burst photons. The evidence for phase-dependence using these measures is mixed (see Section \ref{sec:Overview of published burst phase-dependence analysis} for more details). What has never been done is to determine from a theoretical perspective the circumstances under which bursts from a localized active region would actually give rise to a detectable phase-dependence. This will depend on geometry, gravitational light-bending, any beaming factor associated with the burst emission, the size of the burst sample for a given source, and the intrinsic burst properties (e.g. brightness, duration). Under some circumstances, bursts may well be visible throughout most of the rotational phase cycle even if they do originate from a specific active region. 

In this paper we address this fundamental question of the circumstances under which emission from a localized bursting region would be detectable as a rotational phase dependence according to the measures used in the literature. We then revisit the observational studies carried out to date to see what constraints they actually place on the degree to which bursting might be localized. We also consider the degree to which the rotational phase at which a burst is emitted might affect the properties measured by an observer. 

First, we provide an overview of magnetar burst phase-dependence studies in the literature in Section \ref{sec:Overview of published burst phase-dependence analysis}. Different methods have been applied to various sources in an effort to assess the (non-)phase dependence of magnetar bursts. Next in Section \ref{sec:Methodology}, we briefly outline the method through which we aim to answer the aforementioned questions. We choose to simulate sequences of elementary bursts of which we can control the input parameters and study any phase-dependent effects we may observe. The light curve model is treated in Section \ref{sec:Light curve model} and the simulations are described in Section \ref{sec:Simulations}. We discuss the results of the simulations and assess the claims made in the literature in Sections \ref{sec:Results} and \ref{sec:Discussion and Conclusion}. We find that under certain conditions the properties of the observed bursts may become significantly phase-dependent. However, we also find that for a large range of input burst parameters and configurations, a guaranteed detection of phase-dependence requires many more bursts than have commonly been observed.

\section{Overview of published burst phase-dependence analysis}\label{sec:Overview of published burst phase-dependence analysis}

Here we provide a review of previous work where, given the acquired data, the phase dependence of magnetar bursts has been evaluated. We focus on how the data were obtained and processed, and what method was used to determine the absence or presence of a phase dependence in the burst occurrences or properties. In practice, three methods that have been applied: (i) searching for any significant deviations from uniformity of burst occurrence and photon arrival-time distributions against phase, (ii) searching for any correlation between the phase at which bursts occur and the pulse maxima of the (underlying) pulsed emission, and (iii) Fourier analysis on the burst occurrence times in an effort to search for significant periodicities. The latter has been applied only once; the first two are far more common. It is worth mentioning that the first two methods depend on the accuracy of the ascertained timing ephemeris. The longer the time baseline spanned by the bursts, the greater the risk of undetected time anomalies, such as glitches or spin-down deviations, that may undermine the inference of the phase. Table~\ref{tab:literature review} provides a summary of the references that have carried out phase dependence analysis of magnetar bursts.

\begin{table*}
\caption{Summary of burst phase dependence results in the literature.}
\centering
\begin{tabular}{l|lcccccc}
\hline
Reference & Source & Dates of bursts & \emph{Satellite}/ & $n_{\rm bursts}$ & Method$^{\rm b}$ \\
&& [dd/mm/`yy] & Instrument$^{\rm a}$ &&&& \\
\hline
\hline
\citet{Gavriil2004} & 1E 2259+586 & 18/06/`02 & \emph{RXTE}/PCA & 80 & (i)  \\
\hline
\citet{Savchenko2010} & SGR J1550--5428 & 22/01/`09 & \emph{INTEGRAL}/ACS & 84 & (i)  \\
\citet{Scholz2011} &  & (22/01--30/09)/`09 & \emph{Swift}/XRT & 303 & (i)  \\
\citet{Lin2012} & & 22/01/`09 & \emph{Swift}/XRT & 31 & (i)  \\
&& 30/01/`09 &&&&& \\
\citet{Collazzi2015} & & 03/10/`08--17/04/`09 & \emph{Fermi}/GBM & 354 & (i)  \\
\citet{Mus2015} &  & 22/01/`09 & \emph{RXTE}/PCA & 4 & (ii) \\
&& 06/02/`09 &&&&& \\
&& 30/03/`09 &&&&& \\
&& 11/01/`10 &&&&& \\
\hline
\citet{Gavriil2002} & 1E 1048.1--5937 & 29/10/`01 & \emph{RXTE}/PCA & 2 & (ii)  \\
&& 14/11/`01 &&&&& \\
\citet{Gavriil2006} &   & 29/06/`04 & \emph{RXTE}/PCA & 1 & (ii)  \\
\citet{Dib2009} &  & 29/10/`01 & \emph{RXTE}/PCA & 4 & (ii) \\
&& 14/11/`01 &&&&& \\
&& 29/06/`04 &&&&& \\
&& 28/04/`08 &&&&& \\
\citet{An2014} &  & (17--27)/07/`13 & \emph{NuSTAR} & 8 & (ii)   \\
\hline
\citet{Woods2005} & XTE J1810--197 & 22/07/`03 & \emph{RXTE}/PCA & 6 & (ii)  \\
&& 16/02/`04 &&&&& \\
&& 19/04/`04 &&&&& \\
&& 19/05/`04 &&&&& \\
\hline
\citet{Gavriil2011} & 4U 0142+61 & (06/04--26/06)/`06 & \emph{RXTE}/PCA & 6 & (ii)  \\
\hline
\citet{Palmer1999} & SGR 1806--20 & (10--15)/11/`83 & \emph{ICE} & 33 & (iii)  \\
\hline
\citet{Palmer2002} & SGR 1900+14 & - & - & - & - \\
\hline
\end{tabular}\begin{flushleft}{\small $^{\rm a}$Spacecraft/instrument acronyms: \emph{Rossi X-ray Timing Explorer} (\emph{RXTE}), Proportional Counter Array (PCA), \emph{Nuclear Spectroscopic Telescope Array} (\emph{NuSTAR}), \emph{International Cometary Explorer} (\emph{ICE}), Anti-Coincidence Shield (ACS), X-ray Telescope (XRT), \emph{Nuclear Spectroscopic Telescope Array} (\emph{NuSTAR}), and Gamma-ray Burst Monitor (GBM). \\$^{\rm b}$The methods are specified in Section \ref{sec:Overview of published burst phase-dependence analysis}.}\end{flushleft}
\label{tab:literature review}
\end{table*}

The active phase of 1E 2259+586 on 2002 June 18 consisted of 80 bursts and was studied using method (i)  \citep{Gavriil2004}; it was claimed that the burst peak phase occurrences tended to correlate with the intensity of the pulsed emission, yet no phase dependencies were observed for the burst durations, fluences, peak fluxes, and rise/fall-times.

In excess of 300 bursts were observed from SGR J1550--5428 between 2008 March and 2010 January and many of those bursts were detected by multiple space-based telescopes simultaneously. \citet{Savchenko2010} found that the burst start times (the moment the burst exceeds 5$\sigma$ above background) of 84 bursts, observed with the Anti-Coincidence Shield (ACS) aboard the INTEGRAL spacecraft, appear to be distributed randomly across phase, i.e. no significant departure from the mean bursts per phase bin was identified. \citet{Scholz2011} and \citet{Collazzi2015} studied the burst peak times of, respectively, 303 and 354 bursts, and both found no significant ($>3\sigma$) deviations from the mean number of burst peaks per phase bin. \citet{Scholz2011} however do show that the phase-folded photon times of arrival of the bursts exhibit an apparent pulse which has an offset with respect to maximum of the associated quiescent pulse profile. \citet{Lin2012} study a sample of 31 bursts and similarly find that the burst count distribution is not uniform across phase. Moreover, they find that the phase probability density anti-correlates with the phase profile of the persistent emission (with a correlation factor of -0.5 and chance probability of $3.4\times10^{-2}$), which may suggest that the burst emission region is distinct to that of the persistent emission. Contrary to these results however, \citet{Collazzi2015} do not find a significant ($>3\sigma$) pulse shape in the epoch folded burst emission light curves. Note that the data set used by \citet{Lin2012} constitutes a subset of the data used by \citet{Collazzi2015}.

A total of 12 bursts from 1E 1048.1--5937, were analyzed using method (ii); 4 of which were observed with \emph{RXTE}/PCA between 2001 October 29 and 2008 April 28 \citep{Gavriil2002,Gavriil2006,Dib2009} and 8 of which were observed with \emph{NuSTAR} in 2013 July 17--27 \citep{An2014}. It was determined that the majority of the bursts\footnote{\citet{Dib2009} discuss the 4 bursts from AXP 1E 1048.1--5937 and note that only 3 of them occur near pulse maximum, whereas the fourth burst does not.} observed with \emph{RXTE}/PCA had a probable chance alignment with pulse maxima of less than 0.01. For the latter 8 bursts from the same source however there is no evidence for a preferred phase occurrence. 6 bursts from XTE J1810--197 observed with \emph{RXTE}/PCA were also studied with method (ii) \citep{Woods2005}. These bursts consisted of individual burst spikes, which in turn occurred near the corresponding pulse maxima of the source, either leading or trailing. A chance alignment of these spikes with the pulse maxima was estimated at roughly 0.004. \citet{Gavriil2011} studied 6 bursts of 4U 0142+61 observed with \emph{RXTE}/PCA in 2006 from April to June. They found that several bursts appear to occur near the maxima of contemporaneous folded pulse profiles (no significance criteria are specified in the reference). They argue that this may indicate that the bursts comprise extreme episodes of local transient emission sites.

\citet{Palmer1999,Palmer2002} studied the burst properties of SGR 1806--20 and SGR 1900+14, where for the former source it was found that active bursting episodes emerge from local active regions characterized as `relaxation systems'. From a larger burst sample a group of 33 bursts were identified as belonging to a single relaxation system. Method (iii), i.e. Fourier analysis on the burst occurrences of this group revealed no apparent modulation at the rotation frequency of the NS, indicating the lack of a phase dependence\footnote{No further details of the analysis procedure on the SGR 1806--20, such as the applied nominal threshold to determine significance, are given in the article. The phase dependence analysis procedure of the SGR 1900+14 data is also not described.}.

Here we will focus mainly on the (non-)uniform phase occurrence of burst peaks as a proxy for the phase dependency of magnetar bursts. We briefly discuss the use of alternative methods in Section \ref{sec:Discussion and Conclusion}.

\section{Methodology}\label{sec:Methodology}

To understand how the observed emission may depend on the rotational phase, we intentionally introduce a phase-dependency by fixing the burst location to a certain region or burst patch on the magnetar surface and then set out to describe and simulate the process from emission, where we control the input parameters, to detection and characterization. Subsequently, we can study the effects of a certain configuration on the burst parameters by investigating the phase distributions of the observed burst properties. Moreover, we can establish detectability criteria for the phase-dependency for certain input values/distributions and system configurations.

In order to do so, we require a light curve model that describes how the burst emission is modified depending on the location of the bursts and additional system parameters, e.g. the inclination angle to the observer and compactness of the source. The latter parameter will reshape the trajectory of emitted photons through gravitational light bending.

In Section \ref{sec:Light curve model} we ascertain an expression that describes the fraction of rays, i.e. paths along which the emitted photons propagate, that extend out from the burst location and intersect with an observer at infinity. Subsequently, in Section \ref{sec:Simulations}, we simulate sequences of bursts and investigate how the burst properties are modified through the correction of the burst intensity by the aforementioned expression. 

\section{Light curve model}\label{sec:Light curve model}

\begin{figure*}
	\centering
		\includegraphics[width=\textwidth]{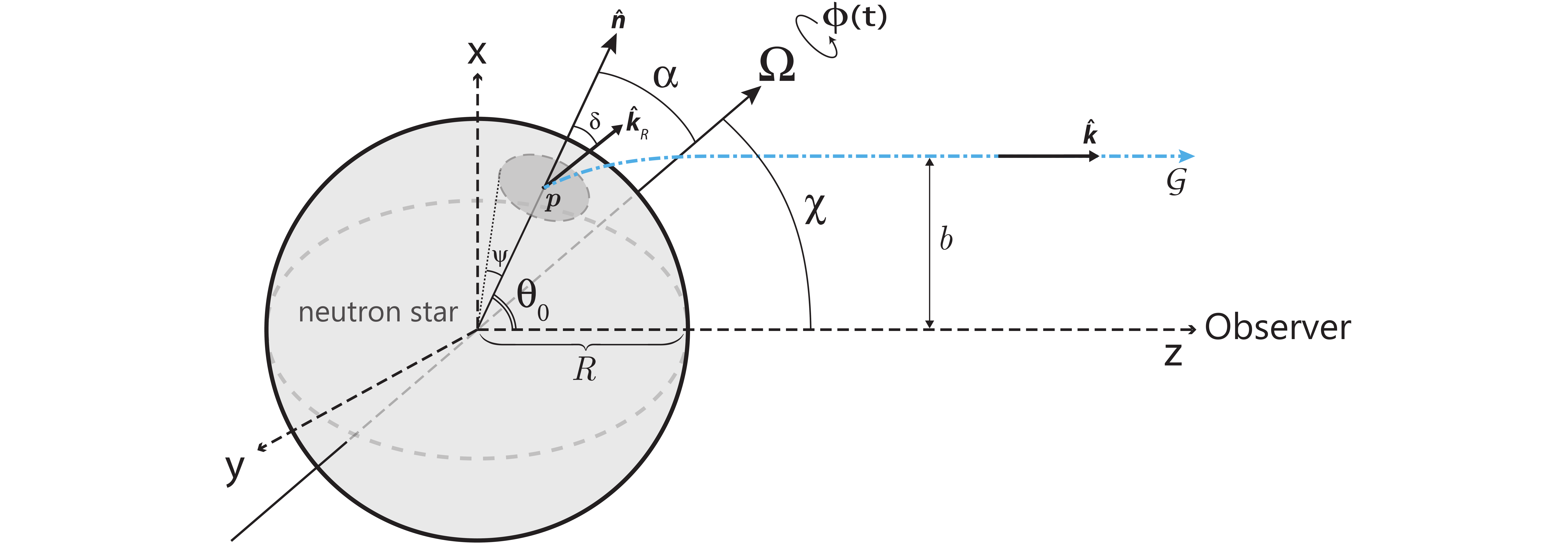}
	\caption{Schematic representation of null geodesic $\mathcal{G}$ connecting the burst patch, centered at point $\boldsymbol{p}(\theta_0)$ of angular width $\psi$, with the observer. The observer is set in the $+z$-direction, where the $z$-axis makes an angle $\chi$ with the rotational axis of the neutron star (NS) $\boldsymbol{\Omega}$. The normal vector to the NS surface $\hat{\boldsymbol{n}}$ at $\boldsymbol{p}$ is at an angle $\alpha$ with $\boldsymbol{\Omega}$ and at an angle $\theta_0$ with the observer's line of sight, where the latter depends on the rotational phase of the NS, i.e. $\theta_0=\theta_0(\phi)$ as prescribed by Eq.~(\ref{eq:cosine theta0}). In a stationary frame at $r = R$, photons radiate from the burst location (along $\hat{\boldsymbol{k}}_R$) at and angle $\delta$ to the normal. The intensity of the source may be isotropic or beamed $I = I(\delta)$ depending on the underlying emission mechanism and properties of the emitting region.}\label{fig:light_bending_configuration}
\end{figure*}

In the following we adopt natural units, i.e. $G=c=1$, and the spatial spherical coordinates ($r,\varphi,\theta$), where $\varphi$ is the polar angle to the $y$-axis and $\theta$ is azimuthal angle to the $z$-axis. Since magnetars rotate slowly (typically $|\boldsymbol{\Omega}|\sim10^{-1}$ rad s$^{-1}$) they can be considered to be almost spherically symmetric. Accordingly, we may assume that the metric external to the star is approximately given by the Schwarzschild spacetime solution,
\begin{equation}
ds^2=-\mathcal{A}(r)dt^2+\mathcal{A}^{-1}(r)dr^2+r^2 \left(d\varphi^2+\sin^2\varphi\,d\theta^2\right)
\end{equation}
where $\mathcal{A}(r)=(1-R_S/r)$, and $R_S=2M$ the Schwarzschild radius, with $M$ corresponding to the gravitational mass of the compact object. To model the effect of gravitational light bending on the burst emission we consider the configuration illustrated in Fig.~\ref{fig:light_bending_configuration}, which is based on work done by \citet{Pechenick1983}. The stellar surface is located at a distance $R$ from the origin; for neutron stars, $R$ lies roughly in the range $2.5-4~R_S$. For now we assume that the burst emission originates at \footnote{Note that here we only consider the case where burst emission escapes from the system at the stellar surface \citep[as it would from a trapped fireball, for example, due to the reduced scattering opacity close to the surface][]{Thompson1995}.  It is for surface emission that the effects of GR will be most significant. We argue that bursts that occur high-up in the magnetosphere will be much less affected by the effects of GR or occultation of the star itself, and thus may exhibit weak to no phase-dependent properties. In effect we are considering the most optimistic case for the detection of phase-dependent effects.} $r = R$ over a circular patch of angular radius $\psi$ centered at point $\boldsymbol{p}(R,\varphi,\theta)$ with total intensity $I$. Depending on the burst emission mechanism, $I$ may be anisotropic and depend on $\delta\in[0,\pi)$, i.e. the angle between the normal vector to the stellar surface $\hat{\boldsymbol{n}}$ and the outgoing emission vector $\hat{\boldsymbol{k}}_R$. The latter lies at $r=R$ along the associated null geodesic $\mathcal{G}$ of the outgoing emission which in turn intersects with the observer at $r=r_0$ where $\hat{\boldsymbol{k}}_R\to\hat{\boldsymbol{k}}$. We presume that the region $r < R$ is opaque and $r > R$ is entirely transparent. Moreover, due to the comparatively long rotation period of magnetars, we may neglect certain corrections, such as oblateness of the stellar surface, light travel-time delays, and Doppler effects, which become significant for NSs with  $|\boldsymbol{\Omega}|\gg1$ rad s$^{-1}$ \citep[e.g.][]{Morsink2007}.

The Schwarzschild solution admits four Killing vectors associated with conserved quantities that emerge from symmetries inherent in the solution for which,
\begin{equation}
K_\mu\dot{x}^\mu=\text{constant},
\end{equation}
where $\dot{x}^{\mu}=dx^{\mu}/d\lambda$, with $\lambda$ is some affine parameter. Considering the orbital motion of a photon with tangent 4-vector $V^\mu=\dot{x}^{\mu}$ in the equatorial plane, i.e. $\varphi = \pi/2$, the Schwarzschild solution admits two Killing vectors
\begin{align}
&\epsilon_\mu=(\partial_t)_\mu=\big(-\mathcal{A}(r),0,0,0\big),\\
&J_\mu=(\partial_\varphi)_\mu=\big(0,0,0,r^2\big),
\end{align}
associated respectively with conservation of energy and the magnitude of angular momentum. Accordingly, we may define
\begin{align}
&\epsilon_\mu\dot{x}^\mu=-\mathcal{A}(r)V^t\equiv-1,\\
&J_\mu\dot{x}^\mu=r^2V^\theta\equiv b,
\end{align}
where $b\geq0$ denotes the impact parameter of the photon trajectory, i.e. the null geodesic $\mathcal{G}$. Since $\dot{x}^\theta=0$ and $g_{\mu\nu} \dot{x}^\mu \dot{x}^\nu=0$ for massless particles, we find that the tangent 4-vector of an outgoing photon is given by,
\begin{equation}
V^{\mu}=\left(\mathcal{A}^{-1}(r),\sqrt{1-\frac{b^2}{r^2}\mathcal{A}(r)},0,\frac{b}{r^2}\right).
\end{equation}
Setting $b=0$, we obtain the tangent 4-vector of a radially outgoing photon,
\begin{equation}
W^{\mu}=\left(\mathcal{A}^{-1}(r),1,0,0\right).
\end{equation}
A stationary observer with 4-velocity
\begin{equation}
U^{\mu}=\left(\mathcal{A}^{-1/2}(r),0,0,0\right),
\end{equation}
will observe an angle 
\begin{equation}\label{eq:observed photon angle}
\cos\xi=\sqrt{1-\frac{b^2}{r^2}\mathcal{A}(r)},
\end{equation}
between the photons prescribed by $V^\mu$ and $W^\mu$ \citep{Pechenick1983}. Note that $\hat{\boldsymbol{n}}\cdot\hat{\boldsymbol{k}}_R=\cos\delta = \cos\,(\xi|_{r=R})$, such that we may write
\begin{equation}
\delta(b)=\arcsin\left(\frac{b}{b_{\rm max}}\right),~~~~\text{with}~~b_{\rm max}=\frac{R}{\mathcal{A}^{1/2}(r=R)}.
\end{equation}
The total angular deflection of $\mathcal{G}$, which determines the `bending' of the photon trajectory from the surface patch to the observer, is given by
\begin{equation}\label{eq:angular deflection}
\theta_*(b) =\int_R^{r_0}\frac{b}{r^2}\left[1-\frac{b^2}{r^2}\mathcal{A}(r)\right]^{-1/2}dr.
\end{equation}
Incidentally, the total coordinate light travel time along $\mathcal{G}$ is given by
\begin{equation}
T_*(b) =\int_R^{r_0}\mathcal{A}^{-1}(r)\left[1-\frac{b^2}{r^2}\mathcal{A}(r)\right]^{-1/2}dr.
\end{equation}

The difference in travel time between radially emitted photons (with $b=0$) and those with an arbitrary impact parameter can be estimated accordingly,  
\begin{equation}
\Delta t_*(b) =\int_R^{r_0}\mathcal{A}^{-1}(r)\left\{\left[1-\frac{b^2}{r^2}\mathcal{A}(r)\right]^{-1/2}-1\right\}dr.
\end{equation}
The maximum travel time delay then for a typical NS with $R = 2.5\,R_S$ ($R = 10^6$ cm, $M = 1.5\,M_\odot$), is $\Delta t_*(b_{\rm max})\simeq6.7\times10^{-2}$ ms $\ll P\sim6$ s.

To an observer in the $+z$-direction the system is axisymmetric around the $z$-axis, such that the location of the burst patch $\boldsymbol{p}$ can be uniquely described by $\theta_0$. The angle between the observer's line of sight and the rotation axis of the neutron star $\boldsymbol{\Omega}$ is denoted by $\chi$. Furthermore, the angle between the location of the burst patch and $\boldsymbol{\Omega}$ is given by $\alpha$. Accordingly, depending on the rotational phase of the neutron star, 
\begin{equation}\label{eq:phase}
\phi(t)=\frac{2\pi t}{P},
\end{equation}
the angle between the observer and burst patch $\theta_0$ is given by the relation
\begin{equation}\label{eq:cosine theta0}
\cos[\theta_0(\phi)] = \cos\chi\cos\alpha + \sin\chi\sin\alpha\cos\phi. 
\end{equation}

The observed brightness is the integral of the intensity at the observer,
\begin{equation}
dI_{\rm obs}=I(r_0,\Omega')d\Omega'=\mathcal{A}^2(r=R)I(R,\Omega')d\Omega',
\end{equation}
over the solid angle subtended in the observer's sky,
\begin{equation}
d\Omega'=\sin\theta' d\theta' d\varphi'\simeq\theta' d\theta' d\varphi'.
\end{equation}
Due to the axisymmetry $d\varphi'=d\varphi\equiv\Phi(\theta_*,\theta_0)$, where we define the polar differential distance as the function $\Phi(\theta_*,\theta_0)$, which under the conditions $\theta_0+\psi\leq\theta_*\leq\pi$ and $\theta_0-\psi\geq0$ is given by the following expression,
\begin{align}
\Phi(\theta_*,\theta_0) = \left\{
\begin{array}{ll}
2\arccos\left(\frac{\cos\psi-\cos\theta_0\cos\theta_*}{\sin\theta_0\sin\theta_*}\right) & \text{if } \theta_0-\psi\leq\theta_*\leq\theta_0+\psi,\\
0 & \text{otherwise.}
\end{array} \right.
\end{align}
Consequently, together with $\theta'=\xi$ and Eq.~(\ref{eq:observed photon angle}) evaluated at $r_0\to\infty$, we obtain 
\begin{equation}
d\Omega'\simeq \Phi[\theta_*(b),\theta_0]\,\xi d\xi\simeq\frac{\Phi(b,\theta_0)}{r_0^2}b\,db.
\end{equation}
We write the brightness of the source as
\begin{equation}
I(R,\Omega')=I_0f[\delta(b)],
\end{equation}
with the beaming functions given by $f(b)$. Currently, we do not have a physical model for the shape of the beaming function, which will most likely depend on the radiative transfer properties of the local magnetic field. Accordingly, the influence of the magnetic field on the light trajectories is ignored for now. An example of a more realistic model was considered by \citet{vanPutten2016} in the case of fireball beaming. Here, for descriptive purposes we consider a Gaussian shape for the beaming function,  
\begin{equation}\label{eq:beaming function}
f[\delta(b)]= \left\{
\begin{array}{ll}
1 & \text{isotropic}, \\
\sqrt{\frac{\pi}{2\sigma_{\rm b}^2}}~{\rm erf}\left(\frac{\pi}{2\sqrt{2}\sigma_{\rm b}}\right)^{-1}\exp\left(-\frac{\delta(b)^2}{2\sigma_{\rm b}^2}\right)& \text{beamed},
\end{array} \right.
\end{equation}
where $\sigma_{\rm b}$ parameterizes the beam width. We neglect rotational aberration of light effects, since the star rotates slowly. Finally, we find the expression for the observed intensity of the source
\begin{equation}
I_{\rm obs}(\theta_0)=I_0\left(\frac{R}{r_0}\right)^2\kappa(\theta_0),
\end{equation}
where we define
\begin{equation}\label{eq:kappa}
\kappa(\theta_0)\equiv\left(\frac{R^{1/2}}{b_{\rm max}}\right)^4\int^{b_{\rm max}}_0f[\delta(b)]\,\Phi(b,\theta_0)\,b\,db.
\end{equation}

\begin{figure}
	\centering
		\includegraphics[width=0.45 \textwidth]{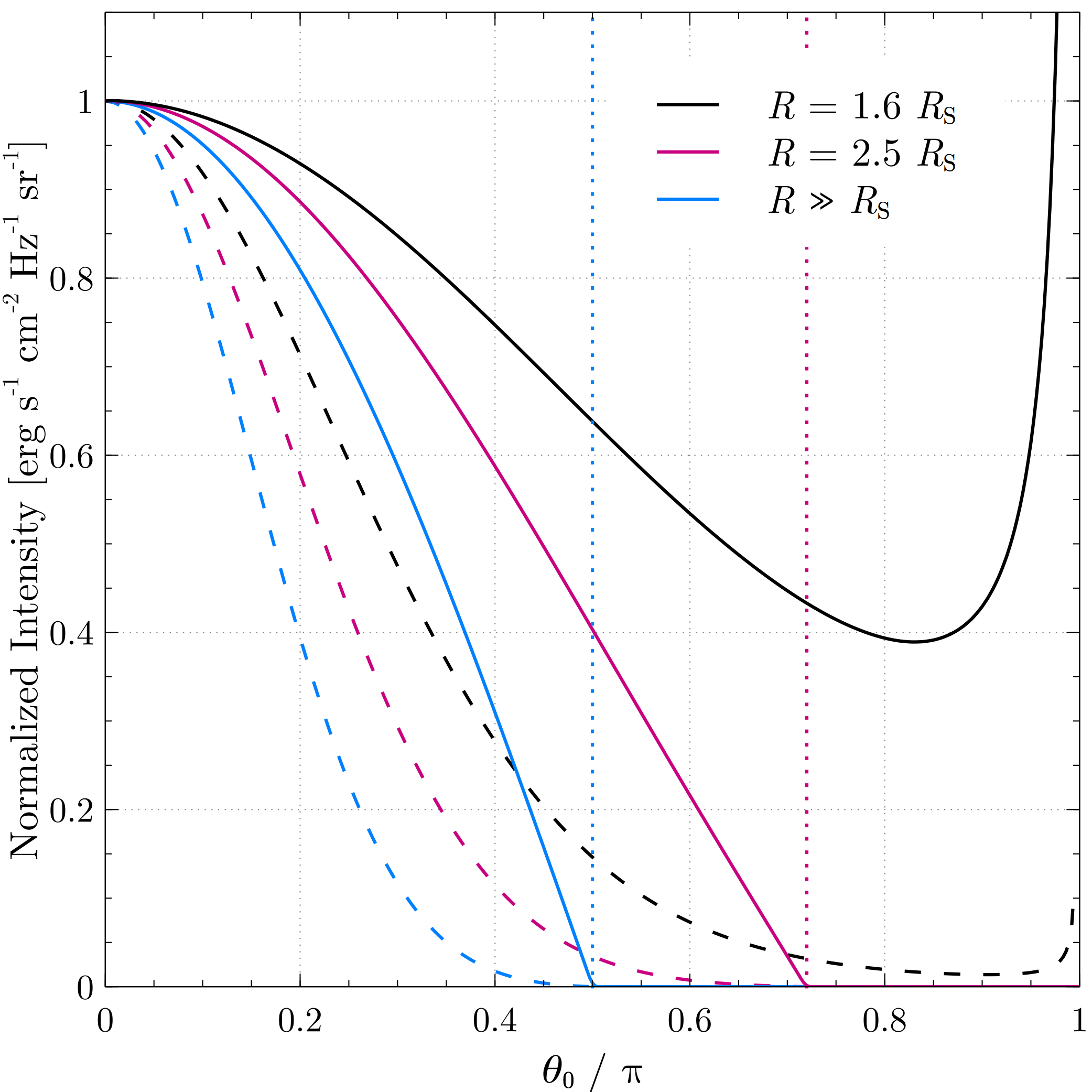}
	\caption{Burst emission as a function of the angle between the observer and the burst location, parameterized by $\theta_0$ (see Fig.~\ref{fig:light_bending_configuration}) for a burst spot area of $\psi=1^\circ$. The solid (dashed) curves denote emission from an isotropic (a beamed) source. A beam width of $\sigma_{\rm b}=\pi/6$ was used. The black curves represent emission from a compact object with $R = 1.6\, R_{\rm S}$ (close to the most extreme case, i.e. $R>1.5\,R_{\rm S}$). The magenta curves denote the emission from a typical neutron star $R = 2.5\, R_{\rm S}$ ($R = 10^6$ cm, $M = 1.5\,M_\odot$). The blue curves denote the emission from a non-relativistic source $R\gg R_{\rm S}$, i.e. if we neglect the GR light bending effects. The dotted lines indicate the location of the terminator, respectively at $\theta_0=0.5\pi$ and $\theta_0\sim0.72\pi$ for $R\gg R_{\rm S}$ and $R = 2.5\, R_{\rm S}$, beyond which the burst patch is invisible. Note that emission is always visible in the $R = 1.6\,R_S$ case regardless the location of the burst, i.e.  $\forall\theta_0$. Moreover, in this extreme case the intensity of the burst patch is greatly amplified at $\theta_0\to\pi$ due to gravitational lensing effects. This plot is similar to Fig. 4 in \citet{Pechenick1983}.}\label{fig:kappa}
\end{figure}

Fig.~\ref{fig:kappa} shows the observed burst emission $I_{\rm obs}$ as a function of burst patch location $\theta_0$ for a compact object with $R=2.5\,R_S$. In this case the size of the patch is $\psi=1^\circ$. We consider the observed intensity for the case of isotropic and beamed emission. Note that the location of the terminator lies `behind' the star, i.e. beyond $\theta_0=\pi/2$, at $\theta_0\simeq0.72\pi$. At this angle, only photons with an impact parameter of $b_{\rm max}\simeq3.23 R_S$ reach the observer. Note that the beamed emission is more prominent at $\theta_0=0$ and drops off faster than the isotropic emission with increasing $\theta_0$. For comparison, we plotted the emission profiles of sources with $R=1.6\,R_S$ and $R\gg R_S$, where the former is close to the most extreme case, i.e. $R>1.5\,R_S$ and the latter approximates flat spacetime.

\begin{figure}
	\centering
		\includegraphics[width=0.45 \textwidth]{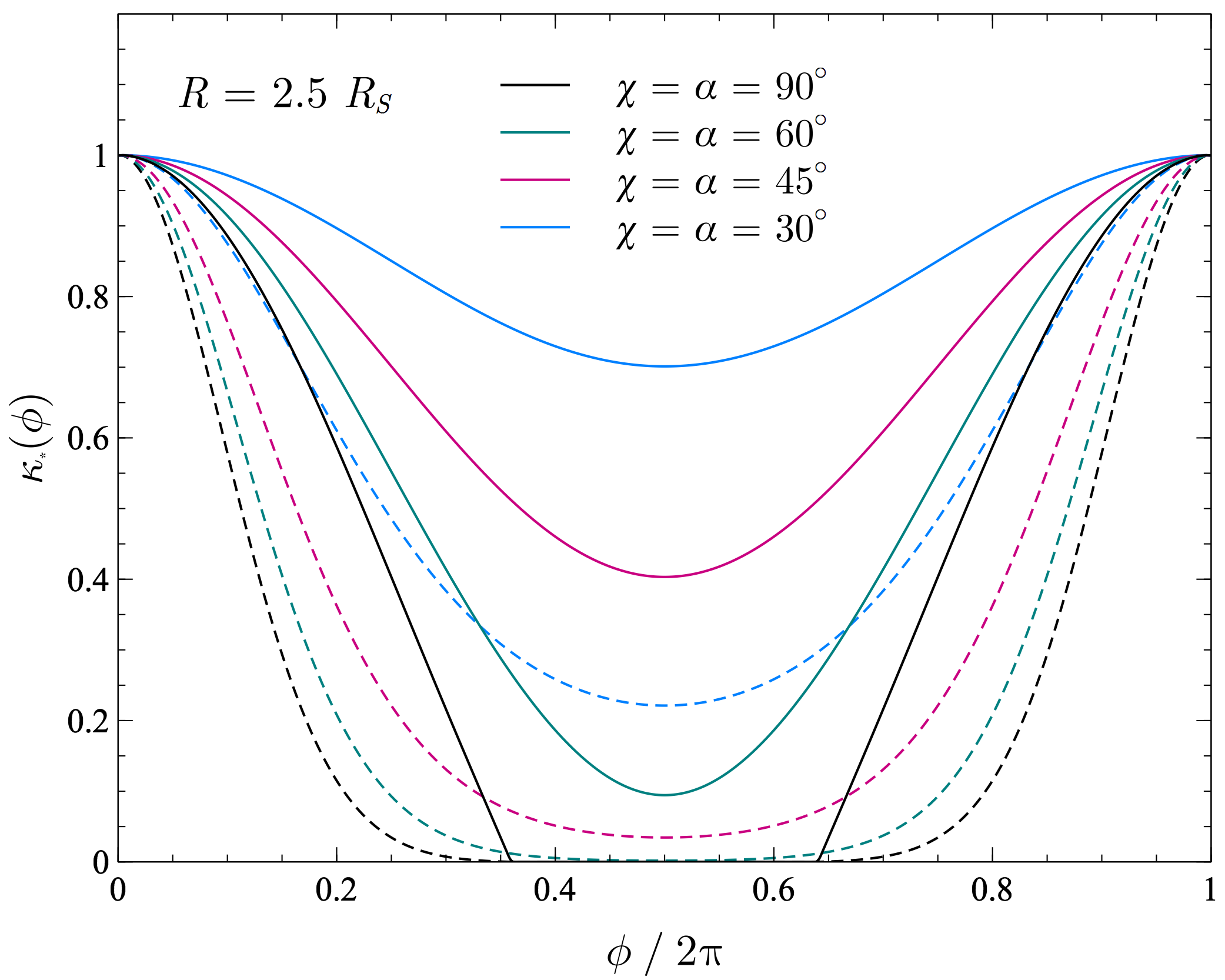}
	\caption{Fraction of rays that intersect with the observer at infinity, from the entire ray-bundle that extends outwards from the burst patch, i.e. $\kappa_*(\phi)$. The solid (dashed) curves represent isotropic (beamed, with $\sigma_{\rm b}=\pi/6$) emission from a source with $R = 2.5\, R_S$. The colours represent 4 different geometries, given by $\chi$ and $\alpha$.}\label{fig:kappa angles}
\end{figure}

In the simulations we concentrate on the relative changes in intensity between the input and observed burst. Accordingly, from equations (\ref{eq:cosine theta0}) and (\ref{eq:kappa}) we define 
\begin{equation}\label{eq:kappa star}
\kappa_*(\phi)\equiv\frac{\kappa[\theta(\phi)]}{\kappa_{\rm max}},
\end{equation}
which depends on the angles $\chi$, $\alpha$, and the phase of the neutron star, and describes the fraction of rays that intersect with the observer at infinity, from the entire ray-bundle that extends outwards from the burst patch. In the following section, we use this expression as our measure for how the burst intensity is modulated. Varying $\chi$ separately from $\alpha$, or vice-versa, acts as a multiplicative factor to the absolute intensity. Since, we only consider the fractional intensity, we may explore the parameter space of these angles by setting $\chi=\alpha$. Fig. \ref{fig:kappa angles} illustrates the shape of $\kappa_*(\phi)$ for $R = 2.5\,R_S$ in 4 different angle configurations, both in the case of isotropic and beamed emission. 

\section{Simulations}\label{sec:Simulations}

\begin{figure}
	\centering
		\includegraphics[width=0.45 \textwidth]{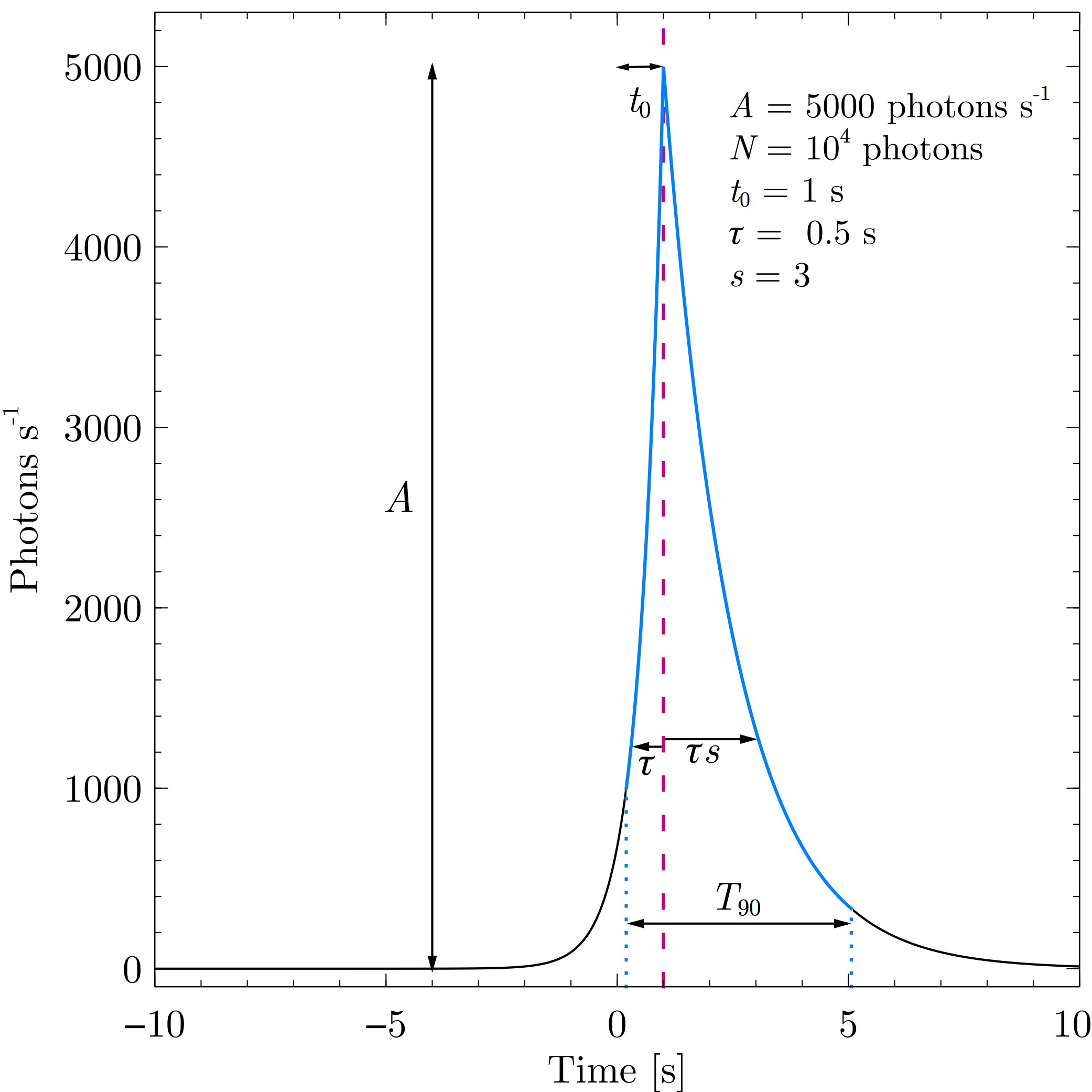}
	\caption{Profile of an input burst, i.e. the photon emission rate at the burst patch, where $t_0$ represents the time at which the burst peaks, $A$ denotes the burst amplitude, $\tau$ denotes the rise-time, $\tau s$ parameterizes the decay-time, determined by $s$ the skewness parameter for a given $\tau$. For $s>1$ ($s<1$) the burst rises faster (slower) than it decays. The burst duration is given by $T_{90}$, which is defined as the time the fluence increases from 0.05 to 0.95 of the total burst fluence; the interval $T_{90}$ contains $0.90 N$ photons.}\label{fig:burst_profile}
\end{figure}

Per simulation run we produce a sequence of $n$ bursts where we control the input parameters of the bursts and the system. Nonetheless, we treat the detection of individual photons, which are entirely described by their times-of-arrival (TOA), in a probabilistic fashion. 

We assume for simplicity that a single magnetar burst can be modeled with a exponential rise/exponential decay profile that allows for asymmetry, i.e. the rise-time and fall-time may be different. When simulating photons emitted at the source, this corresponds to drawing $N$ random photon times-of-emission (TOE) from a skewed Laplace distribution, 
\begin{equation}\label{eq:skewed Laplace probability density function}
p_{\rm TOE}(t)=\frac{1}{(1+s)\tau}\left\{
\begin{array}{ll}
\exp\left[(t-t_0)/\tau\right] & \text{if }~ t < t_0,  \\
\exp\left[-(t-t_0)/(s\tau)\right] &  \text{if }~ t \geq t_0, 
\end{array} \right.
\end{equation}
where $\tau$ denotes the exponential rise-time of the burst, ($s\tau$) the decay-time, with $s$ the skewness factor, and $t_0$ the peak time of the burst. The profile of a simulated burst is shown in Fig.~\ref{fig:burst_profile}. We deliberately adopt an oversimplified burst profile to better understand the differences between the input and output data. \citet{Huppenkothen2015} decompose complex magnetar bursts, observed from SGR J1550-5418, into several spike-like components, which in turn are modeled with a similar profile as in equation (\ref{eq:skewed Laplace probability density function}). In this paper we assume that a burst can be represented as a single spike, as a simple model that lets us explore the relevant effects.  Note that we define the duration of the burst, $T_{90}$, as the time it takes for the fluence to increase from 0.05 to 0.95 of the total burst fluence. We fix the compactness of the source to $R=2.5\, R_{\rm S}$, corresponding to a typical neutron star with $R=10^6$ cm and $M = 1.5 M_\odot$, and set the rotation period to $P=6$ s. We choose a light curve bin width, $\delta t$, and background count level, $b$, such that the background count rate approximates that from \emph{Fermi}/GBM data \citep{Huppenkothen2015}, i.e. $\zeta_{\rm GBM}\sim318$ counts s$^{-1}$.  A list of the simulation parameters is given in Table \ref{tab:sim_pars}.

\subsection{General simulation procedure}\label{sec:General simulation procedure}

Here we proceed to describe in more detail the general form of a simulation run step-by-step:

\begin{enumerate}[(I)]

\item We decide on the number of bursts $n$ we wish to produce per simulation run and set the inclination angle of the source $\chi$. Next, we assign values to the burst parameters $\psi$, $\alpha$, $N$, $t_0$, $\tau$, $s$, and $T_{90}$, where the latter three parameters cannot be defined independently of each other. In the following simulation runs, described in Sections \ref{sec:Run 1}, \ref{sec:Run 2}, and \ref{sec:Run 3}, we only consider symmetric input bursts, $s=1$, with a size of $\psi=1^{\circ}$, and draw their peak time from a uniform distribution, i.e. $t_0\sim{\rm Uniform}(0,P)$, where $P$ is the rotation period of the magnetar which we set to $P=6$.

\item We generate a single burst by drawing $N$ photon emission times (TOEs) from $p_{\rm TOE}(t)$: We draw random numbers from a uniform distribution ${\rm Uniform}(0,1)$ and transform these values to follow the required skewed Laplace distribution by using the inverse cumulative distribution function of the latter, i.e. the percent point function, 
\begin{equation}\label{eq:skewed Laplace percent point function}
{\rm TOE}(x)=\left\{
\begin{array}{ll}
t_0 + \tau\ln\left[\left(1+s\right)x\right] & \text{if }~ x < (1+s)^{-1},\\
t_0 - s\tau\ln\left[\left(1+s^{-1}\right)\left(1-x\right)\right] &  \text{if }~ x \geq (1+s)^{-1}. 
\end{array} \right.
\end{equation}

\item Using equation~(\ref{eq:phase}) we determine the phase of each TOE$_{i}$, i.e. $\phi_i$. Whether an emitted photon reaches the observer depends on whether the ray, along which the photon propagates, intersects with the detector, given by $\kappa_*(\phi_i)$, which denotes the fraction of photons directed into our line of sight. In order to decide whether a given photon intersects with the detector, we use rejection sampling: for each TOE$_{i}$ we generate a latent variable $z$ drawn from $p(z) = {\rm Uniform}(0,1)$. We only keep the TOE$_{i}$ if $z < \kappa_*(\phi_i)$.\footnote{If no burst photons are detected, i.e. $N_{\rm det}=0$, we move on to the next burst [step (II)].} 

\item The TOEs that we save are detected by the observer. A detected photon is recorded as a count with a corresponding TOA$_{i}$; where the TOA$_{i}$ = TOE$_{i}$ of the respective photon, since we consider a perfect detector and may neglect the distance to the source and gravitational time-delay effects -- see Section \ref{sec:Light curve model}. Furthermore, we add background counts or TOAs uniformly to the detected TOA data from the burst (with length $N_{\rm det}\leq N$), such that the mean background count rate becomes approximately $\zeta$.

\item We bin the total TOA data in $\mathcal{N}_{\rm bins}$ time bins of length $\delta t$, whereby the counts in each bin follow Poisson statistics. We proceed by applying a similar burst identification algorithm as used by \cite{Gavriil2004}, assuming that we can infer the background count rate to be $\zeta$, yet have no prior knowledge of the burst and system input parameters. The probability of the number of counts $k_i$ in the $i$th bin occurring is given by the Poisson distribution,
\begin{equation}\label{eq:Poisson distribution}
P_i=\frac{\mu^{k_i}e^{-\mu}}{k_i!},
\end{equation} 
where $\mu$ represents the mean count level, which in our case will be $b=\zeta\,\delta t$. Bins for which 
\begin{equation}\label{eq:burst detection threshold}
P_i \leq 3\times10^{-3}\mathcal{N}_{\rm bins}^{-1}
\end{equation}
are recorded as significant departures from the mean, where we have corrected for the number of trails by dividing by $\mathcal{N}_{\rm bins}$, i.e. the total number of bins searched over. From these, the time bin containing the maximum departure $y^{\rm sig}$ is labeled as $t_0^{\rm sig}$. The burst edges, labeled as $t_{\rm in}$ and $t_{\rm out}$, are ascertained by making use of a running mean, i.e. when the mean count level of an interval $\mu^*$ of $\Delta T_{\rm interval}=0.25$ s, moving outwards in steps of $\delta t$ on both sides of $t_0^{\rm sig}$, falls below $b^*=1.1b$, the burst edges are then given by the center time of the respective intervals; $t_{\rm in}$ before and $t_{\rm out}$ after the burst. The duration of this interval is denoted $\Delta T=|t_{\rm in}-t_{\rm out}|$.

\item We fit the light curve of any identified bursts with the following burst model,
\begin{equation}\label{eq:burst model}
m(t) = N p_{\rm TOE}(t\,|\,t_0,\tau,s) + \zeta,
\end{equation}
using a \emph{L-BFGS-B} constrained optimizer \citep{Zhu1997} to determine the maximum (Poisson) likelihood, whereby we fix the background parameter $\zeta$ and provide initial guesses for the remaining parameters:
\begin{align}\label{eq:initial parameters}
t_0^{\rm init} &= t_0^{\rm sig},\\
\tau^{\rm init} &= \left(t_0^{\rm sig} - t_{\rm in}\right)\left[\ln\left(\frac{y^{\rm sig}}{b^*}\right)\right]^{-1} ,\\
s^{\rm init} &= 1,
\end{align}
and $N^{\rm init}$ is defined as the number of counts in the interval $\Delta T$ minus the background counts\footnote{If $N^{\rm init}<y_0^{\rm sig}$, we set $N^{\rm init}=(1+s^{\rm init})\tau^{\rm init}y_0^{\rm sig}/\delta t$.}, i.e. $\zeta\Delta T$. After the fit, we delete the bins in $\Delta T$ from the light curve, and repeat steps (V) and (VI) until no significant departures from the mean, i.e. $\mu=b$, are recorded. 

\item We return to step (II) until we have generated the pre-defined number of bursts $n$. Note that the number of observed bursts might be different, since bursts might go undetected, or be interpreted as multiple separate bursts. Moreover, some identified `bursts' may simply be significant statistical deviations from the background level. However, according to the condition stated in equation (\ref{eq:burst detection threshold}) we only expect this to be the case in $\sim 0.3$\% of the input bursts.

\end{enumerate}

\subsection{Run 1: Initial simulation run}\label{sec:Run 1}

\begin{figure}
	\centering
		\includegraphics[width=0.45 \textwidth]{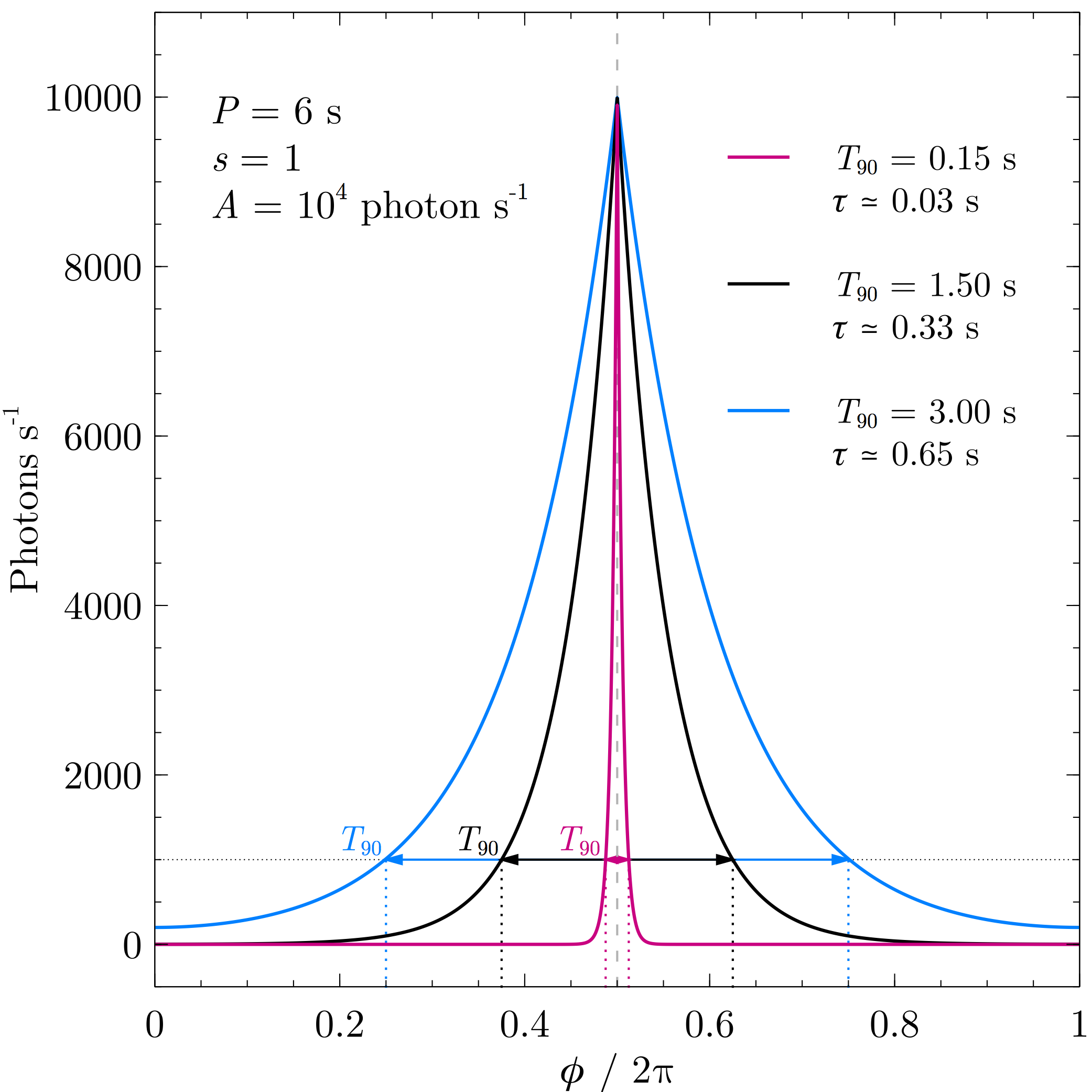}
	\caption{Folded profiles of input bursts of the initial simulation run. We consider symmetric bursts of three separate durations. Here the rise-time $\tau$ is determined by the values for $s$ and $T_{90}$. We choose $N$ such that the burst amplitude is $A=10^4$ photons s$^{-1}$ at $t_0$.}\label{fig:bursts}
\end{figure}

We start with the simplest scenario, where we consider simulations of sequences of identical bursts, referred to as Run 1. Per simulation we fix the values for $\chi$, $\alpha$, $\psi$, $s$ and $T_{90}$. The latter two parameters determine the value of $\tau$. Subsequently, we define $N$ using the condition that the input burst amplitude $A$ is $10^4$ photons s$^{-1}$. We run simulations for three separate burst durations (see Fig~\ref{fig:bursts}) and vary the angles $\chi$ and $\alpha$, to study their effects on the observed quantities. 

We concentrate on the difference in input and observed best-fit value for the time of the burst peak (respectively, $t_0$ and $t_0^{\rm bf}$), where the difference is parameterized as $\Delta t_0\equiv t_0^{\rm bf}-t_0$, the rise-time $\tau$, skewness factor $s$, and burst duration $T_{90}$. The input values of the latter three parameters are denoted as $\tau_0$, $s_0$, and $T_{90,0}$. All input parameters of Run 1 are listed in Table~\ref{tab:RUN1}.  

\begin{table}
\caption{Input parameters for Run 1, consisting of 12 separate simulations. We consider a constant input burst profile with peak times distributed uniformly across phase. Per simulation we vary the burst duration $T_{90}$, and angles $\chi$, $\alpha$; where we set $\chi=\alpha$ (see Section \ref{sec:Light curve model}).}
\centering
\begin{tabular}{ll}
\hline
Parameter & Value \\
\hline
\hline
$n$ & $10^4$ \\ 
$\chi$, $\alpha$ ($^\circ$)& 30, 45, 60, 90\\
$A$ (photons s$^{-1}$) & $10^4$ \\
$T_{90}$ (s) & 0.15, 1.5, 3.0\\
$\delta t$ (s) & $200^{-1}$ \\
\hline
\end{tabular}
\label{tab:RUN1}
\end{table}

\subsection{Run 2: $T_{90}$-distribution}\label{sec:Run 2}

Next we perform a simulation run, Run 2, where in step (I) of the general simulation procedure (Section \ref{sec:General simulation procedure}), we draw the burst duration $T_{90}$ for each individual burst from a lognormal distribution centred at $\overline{T}_{90}=0.1$ s, with a width of $\sigma_{T_{90}}=1$ \citep[e.g.][]{Gogus2001}, and lower and upper burst duration cutoff at, respectively, $T_{90}^{\rm min}=300^{-1}$ s and $T_{90}^{\rm max}=3$ s $< P$. Fixing the input burst amplitude at $A=10^4$ photons s$^{-1}$, as done in Run 1, we find that the rise-time of the shortest admissible burst duration, i.e. $300^{-1}$ s, is $\tau^{\rm min}\sim7.2\times10^{-4}$ s. Accordingly, we set $\delta t = 1400^{-1}$ s for this simulation run. The input parameters are summarised in Table \ref{tab:RUN2} and the results are presented in Section \ref{subsec:Run 2}.

\begin{table}
\caption{Input parameters for Run 2, consisting of 4 separate simulations. The input burst durations $T_{90}$ are drawn from a lognormal distribution with lower and upper cutoff, respectively, at $T^{\rm min}_{90}=300^{-1}$ s and $T^{\rm max}_{90}=3$ s. Per simulation run we vary the angles $\chi$, $\alpha$; where we set $\chi=\alpha$.}
\centering
\begin{tabular}{ll}
\hline
Parameter & Value \\
\hline
\hline
$n$ & $10^4$ \\ 
$\chi$, $\alpha$ ($^\circ$)& 30, 45, 60, 90\\
$A$ (photons s$^{-1}$) & $10^4$ \\
$T_{90}$ (s) & $\sim{\rm LogNormal(\overline{\emph{T}}_{90},\sigma^2_{\emph{T}_{90}})}$\\
$\overline{T}_{90}$ (s) & $0.1$\\
$\sigma_{T_{90}}$ (s) & $1$\\
$\delta t$ (s) & $1400^{-1}$ \\
\hline
\end{tabular}
\label{tab:RUN2}
\end{table}

\subsection{Run 3: Burst amplitude distribution}\label{sec:Run 3}

\begin{table}
\caption{Input parameters for Run 3, consisting of 4 separate simulations. The input burst amplitudes $A$ are drawn from a powerlaw distribution [equation (\ref{eq:burst energy distribution})], with $A^{\rm min}=5\times10^2$ photons s$^{-1}$, and $A^{\rm max}=10^6$ photons s$^{-1}$. Per simulation run we vary the angles $\chi$, $\alpha$; where we set $\chi=\alpha$.}
\centering
\begin{tabular}{ll}
\hline
Parameter & Value \\
\hline
\hline
$n$ & $10^4$ \\ 
$\chi$, $\alpha$ ($^\circ$)& 30, 45, 60, 90\\
$A$ (photons s$^{-1}$) & $\sim{\rm Powerlaw(\Gamma)}$ \\
$\Gamma$  & $5/3$ \\
$T_{90}$ (s) & 1\\
$\delta t$ (s) & $200^{-1}$ \\
\hline
\end{tabular}
\label{tab:RUN3}
\end{table}

In this simulation run we fix the burst duration to $T_{90} = 1$ s and draw an amplitude $A$ for each individual burst from a powerlaw distribution,  
\begin{equation}\label{eq:burst energy distribution}
\frac{dn}{dA}\propto A^{-\Gamma},~~~\text{with}~~~A^{\rm min} < A < A^{\rm max},
\end{equation}
where $A^{\rm min}$ and $A^{\rm max}$ represent the limits of the distribution, and $\Gamma$ denotes the powerlaw index. Note that the number of emitted photons during the burst $N$ are linearly proportional to $A$, such that these are distributed in a similar fashion. 

In accordance with the observation of the energy distributions of magnetar bursts, we choose $\Gamma = 5/3$ \citep[e.g.][]{Cheng1996}. The input parameters are summarised in Table \ref{tab:RUN3} and the results are presented in Section \ref{subsec:Run 3}.

\section{Results}\label{sec:Results}

\subsection{Predictions for Run 1}

\begin{figure*}
	\centering
		\includegraphics[width=0.8\textwidth]{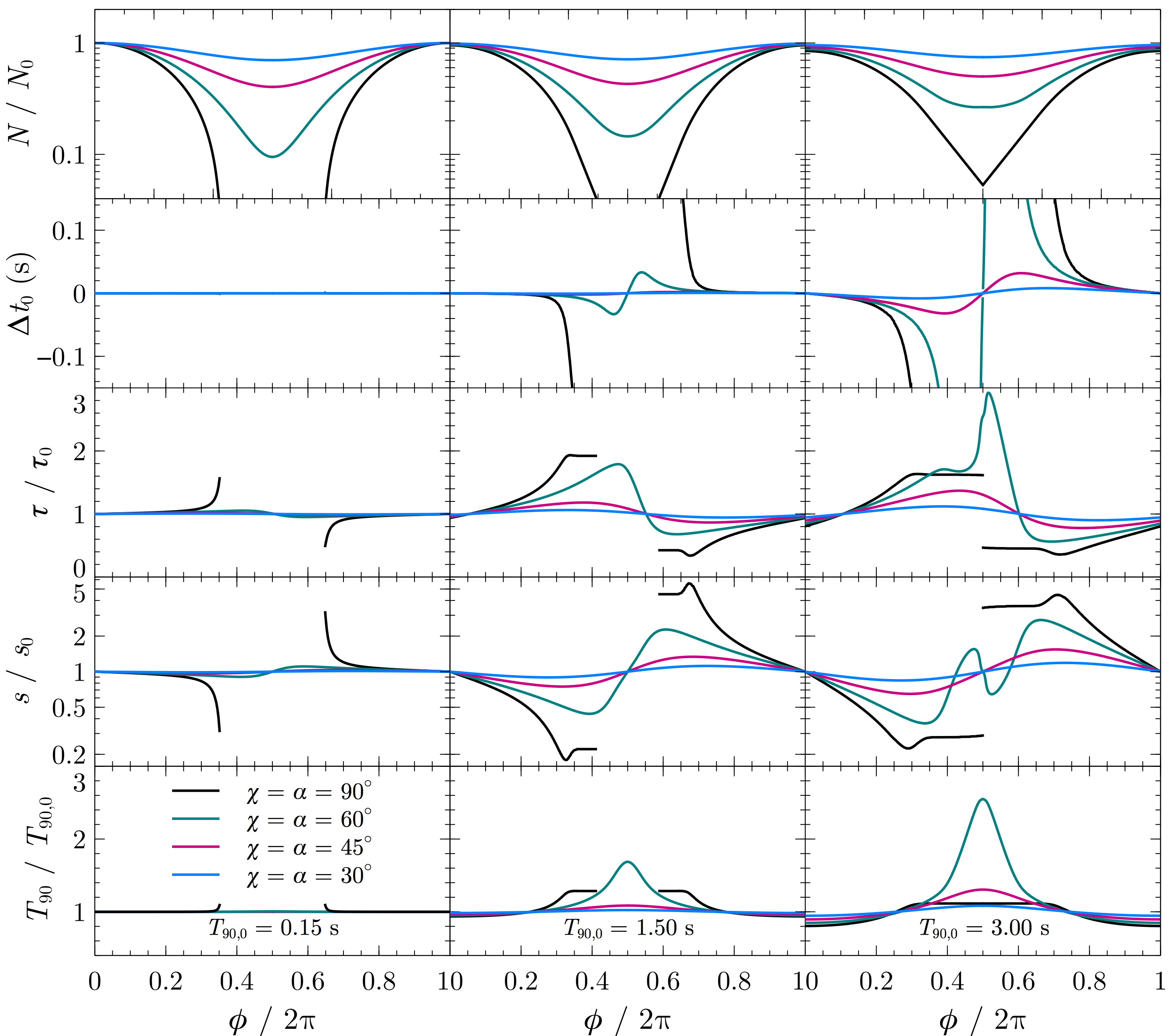}
	\caption{Predicted phase distributions of burst parameters (top to bottom) for the 3 bursts (left to right) studied in the initial simulation run, Run1. From top to bottom: best-fit burst counts, time difference between burst input $t_0$ and the best-fit value ($\Delta t_0$), best-fit burst rise-time $\tau$, best-fit skewness factor $s$, and burst duration $T_{90}$, inferred from the latter $\tau$ and $s$ (the input parameters are given by $N_0$, $\tau_0$, $s_0$,and $T_{90,0}$). The distinct colours represent different values for the angles, $\chi$ and $\alpha$. These curves were obtained by fitting the burst model to theoretical burst profiles (Fig. \ref{fig:bursts}) that are modulated by $\kappa_*$ (equation \ref{eq:kappa star}, Fig. \ref{fig:kappa angles}), depending on their phase occurrence. We only proceed to fit the modulated profile if the peak rate is $\gtrsim600$ counts s$^{-1}$. Note that for longer burst durations and larger angles, the best-fit parameters deviate more from their input values.}\label{fig:RUN1_analytic}
\end{figure*}

To better understand the results of the simulations, we first examine how $\kappa_*(\phi)$ will affect the burst parameters. In Fig. \ref{fig:RUN1_analytic}, we plot the predicted phase distributions of the burst parameters (rows) for the 3 separate burst durations (columns) of Run 1. These curves were obtained by fitting  the burst model to the theoretical lightcurve that results when the input model (see Fig. \ref{fig:bursts}) is modulated, for a given phase, by the appropriate $\kappa_*(\phi)$ but without taking into account any photon noise or detectability effects (which are treated properly in the full simulations). It gives an idea of the general trends expected, but no idea of the scatter. Furthermore, we only fit modulated burst profiles with a peak rate of $\gtrsim600$ counts s$^{-1}$, since ones with lower peak rates will likely go undetected in the simulations.

Based on the predicted curves we expect that the parameter distributions that we obtain from Run 1 will deviate from their input parameters more strongly for longer burst durations $T_{90}$ and larger angles $\chi$ and $\alpha$. Approaching $\phi = \pi$ from below (above), we find that the bursts will appear to occur earlier (later), rise slower (faster), to become more skewed, and last longer, than their input counterparts. Note furthermore that, in contrast to the predicted phase distributions of $N$, $\Delta t_0$, $s$, and $T_{90}$, the phase distribution of $\tau$ is neither symmetric nor perfectly anti-symmetric about $\phi = \pi$. The results of Run 1 are presented in Section \ref{subsec:Run 1}.

\subsection{Burst properties from simulations}

\subsubsection{Run 1: Initial simulation run}\label{subsec:Run 1}

\begin{figure}
	\centering
		\includegraphics[width=0.45 \textwidth]{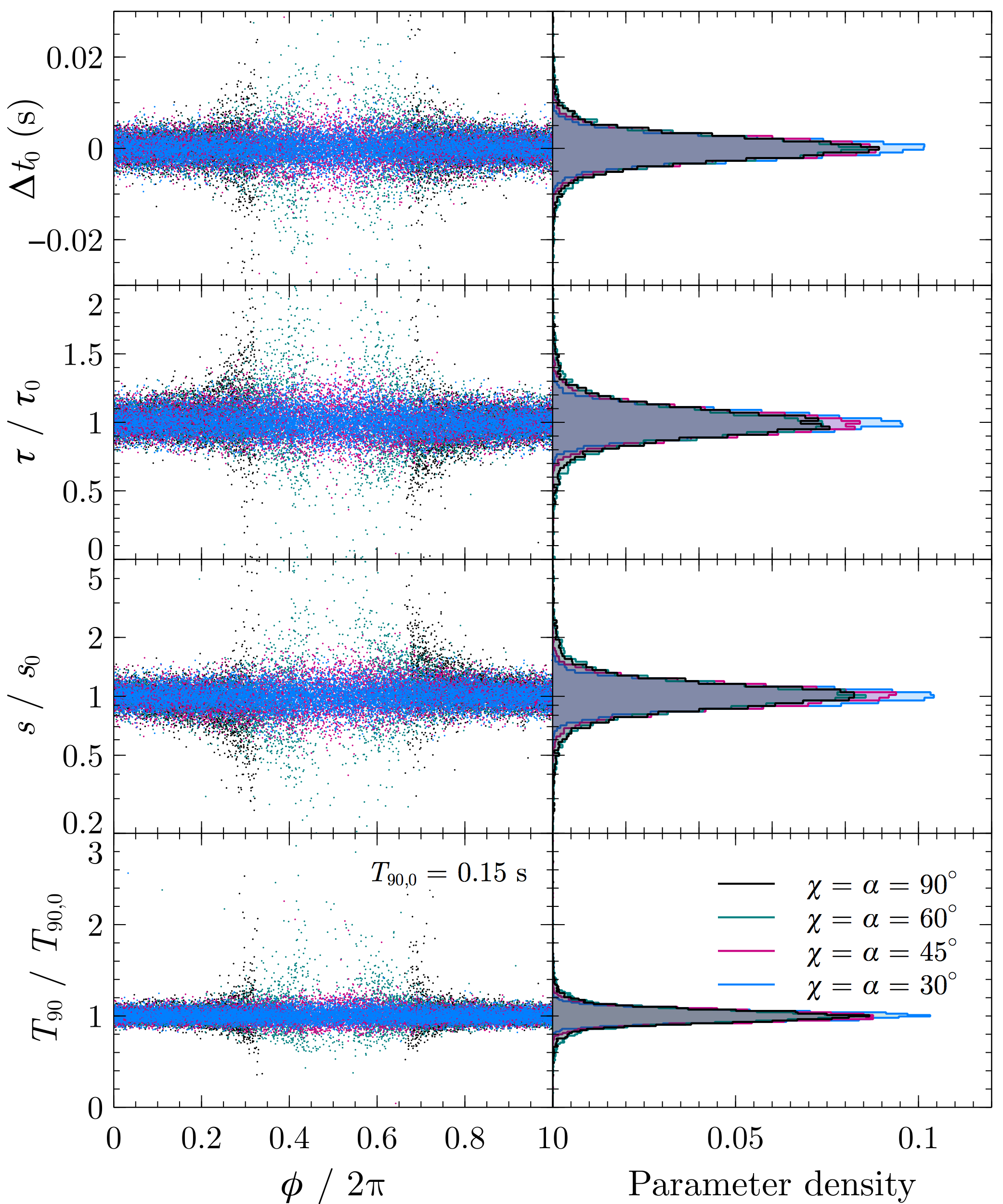}
	\caption{Phase distributions (left) and parameter densities (right) of burst parameters of Run 1 for an input burst of $T_{90,0}=0.15$ s. The theoretical predictions for these distributions are shown in Fig. \ref{fig:RUN1_analytic}.}\label{fig:RUN1_0.15}
\end{figure}

\begin{figure}
	\centering
		\includegraphics[width=0.45 \textwidth]{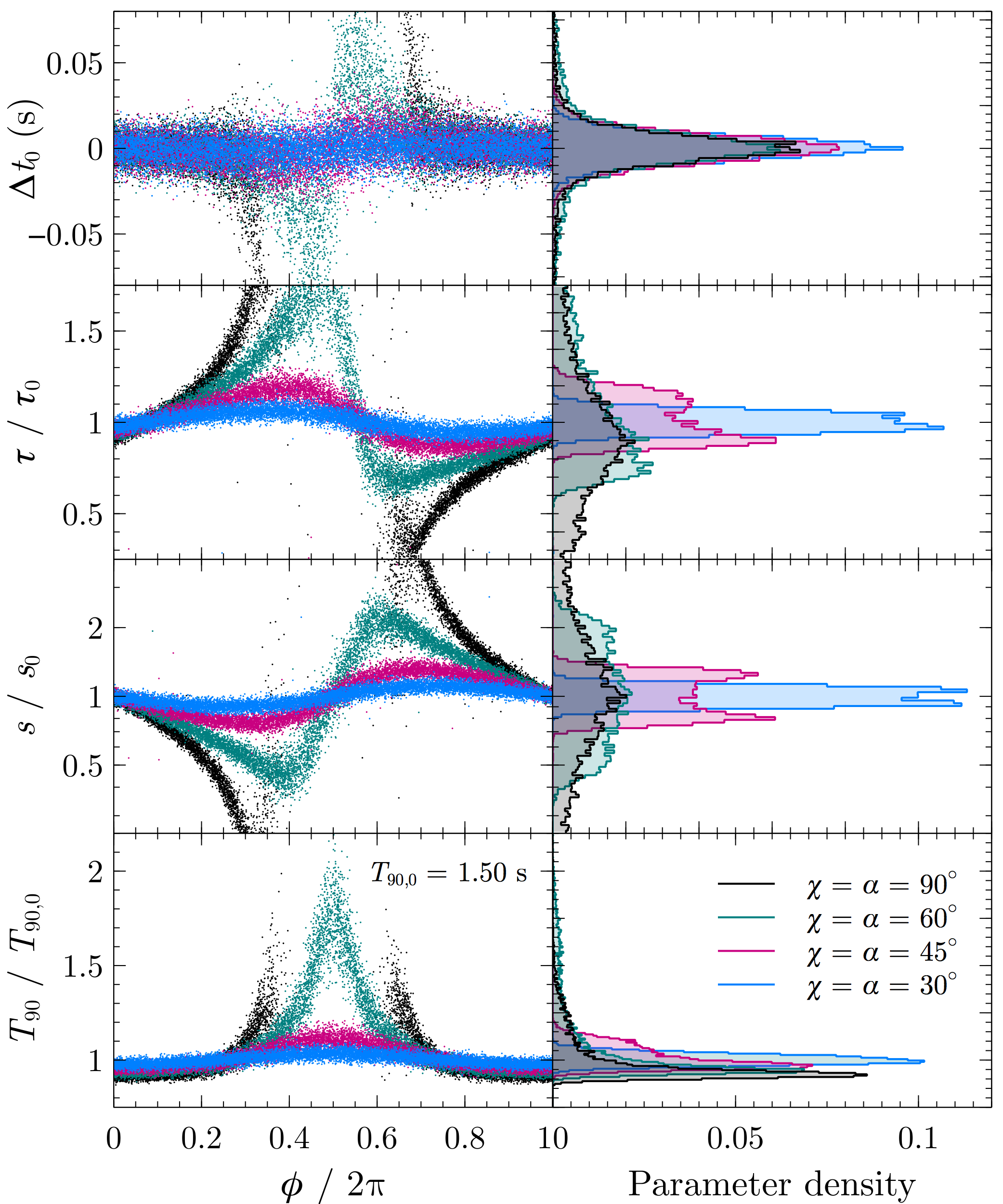}
	\caption{Phase distributions (left) and parameter densities (right) of burst parameters of Run 1 for an input burst of $T_{90,0}=1.5$ s. The theoretical predictions for these distributions are shown in Fig. \ref{fig:RUN1_analytic}.}\label{fig:RUN1_1.50}
\end{figure}

\begin{figure}
	\centering
		\includegraphics[width=0.45 \textwidth]{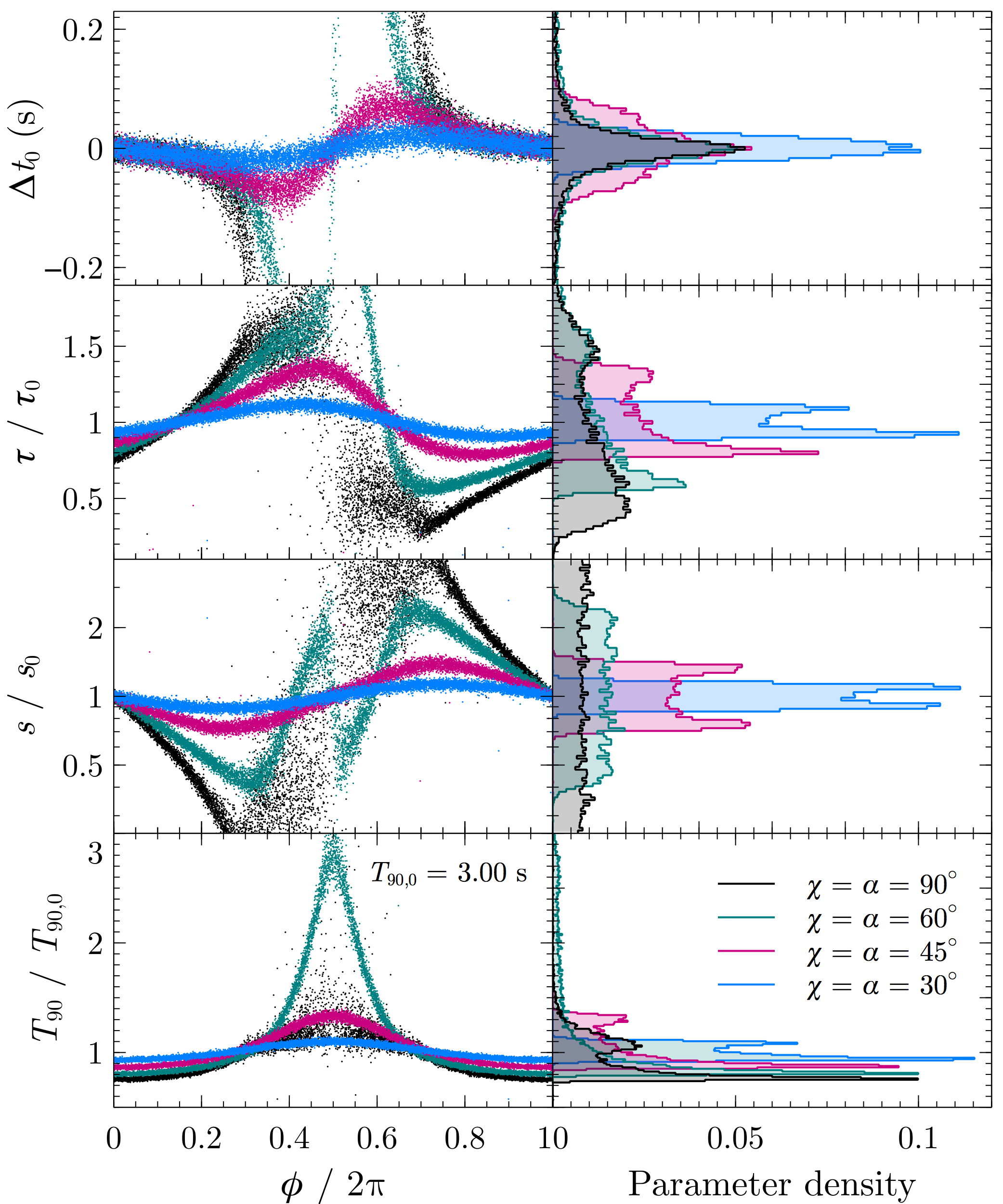}
	\caption{Phase distributions (left) and parameter densities (right) of burst parameters of Run 1 for an input burst of $T_{90,0}=3.0$ s. The theoretical predictions for these distributions are shown in Fig. \ref{fig:RUN1_analytic}.}\label{fig:RUN1_3.00}
\end{figure}

In Figures \ref{fig:RUN1_0.15}, \ref{fig:RUN1_1.50}, and \ref{fig:RUN1_3.00}, we plot the phase distributions (left) and parameter densities (right) of the obtained bursts parameters for 3 separate input burst durations $T_{90,0}$, respectively, 0.15 s, 1.5 s, and 3.0 s. Table \ref{tab:sig_bursts_RUN1} lists the amount of bursts that were identified per configuration.

\begin{table}
\caption{Number of identified bursts $n_{\rm id}$ for Run 1, per configuration. The input number of bursts for each simulation run was $n=10^4$. We expect that $\sim30$ of the identified `bursts' simply constitute statistical deviations that exceed the burst identification threshold (given by equation \ref{eq:burst detection threshold}).}
\centering
\begin{tabular}{lcc}
\hline
$T_{\rm 90}$ (s) & $\chi=\alpha$ ($^\circ$) & $n_{\rm id}$ \\
\hline
\hline
0.15 & 30 & 10034 \\
& 45 & 10022 \\
& 60 & 8875 \\
& 90 & 6592 \\
\hline
1.50 & 30 & 10012 \\
& 45 & 10017\\
& 60 & 9940 \\
& 30 & 7438 \\
\hline
3.00 & 30 & 10017 \\
& 45 & 10012 \\
& 60 & 10007 \\
& 90 & 9553 \\
\hline
\end{tabular}
\label{tab:sig_bursts_RUN1}
\end{table}

As predicted, we find that especially for longer duration bursts and larger angles, the phase dependence of the burst parameters becomes more pronounced. Evidently this is much less the case for bursts with $T_{90}\ll P$ -- the parameter densities remain strongly peaked around their input values (e.g. Fig.~\ref{fig:RUN1_0.15}). Nevertheless, in those cases around $\phi\sim\pi$ the bursts still go undetected for large values of $\chi$ and $\alpha$, because either no rays extending from the burst patch intersect with the detector during the burst (i.e. the bursts are invisible) or they do not significantly stand out from the background level. The results confirm that when approaching $\phi = \pi$ from below (above), the bursts will appear to occur earlier (later), rise slower (faster), and last longer, than their input counterparts. Moreover, the predicted asymmetric profile of the rise-time phase distribution (most notably in the parameter densities of Figures \ref{fig:RUN1_1.50} and \ref{fig:RUN1_3.00}) is clearly observed. Looking at the parameter densities, it appears that the rise-times of bursts going out of view are more spread out, yet those of bursts coming into view are more clustered. This is in accordance with the predictions; the initial slope of the rise-time phase distribution (from $\sim0-4\pi/5$) is steeper compared to the final slope (from $\sim6\pi/5-2\pi$). The predicted values between $\sim4\pi/5-6\pi/5$ are produced less well in the simulations, since the amount of detected photons is minimal around $\phi = \pi$, complicating burst-identification and characterization. We find that both in the predictions and, even more so, in the simulations that the majority of observed bursts have $\tau/\tau_0<1$. Since, $T_{90}\propto\tau$ we also find for most observed bursts that $T_{90}/T_{90,0}<1$, i.e. the bursts seem to last shorter than their input counterparts.

In general, the observed scatter is likely due to photon noise effects, which become most significant near $\phi=\pi$. These effects influence the efficacy of the burst-identification algorithm and the observed burst morphology.

\subsubsection{Run 2: $T_{90}$ distribution}\label{subsec:Run 2}

\begin{figure}
	\centering
		\includegraphics[width=0.45 \textwidth]{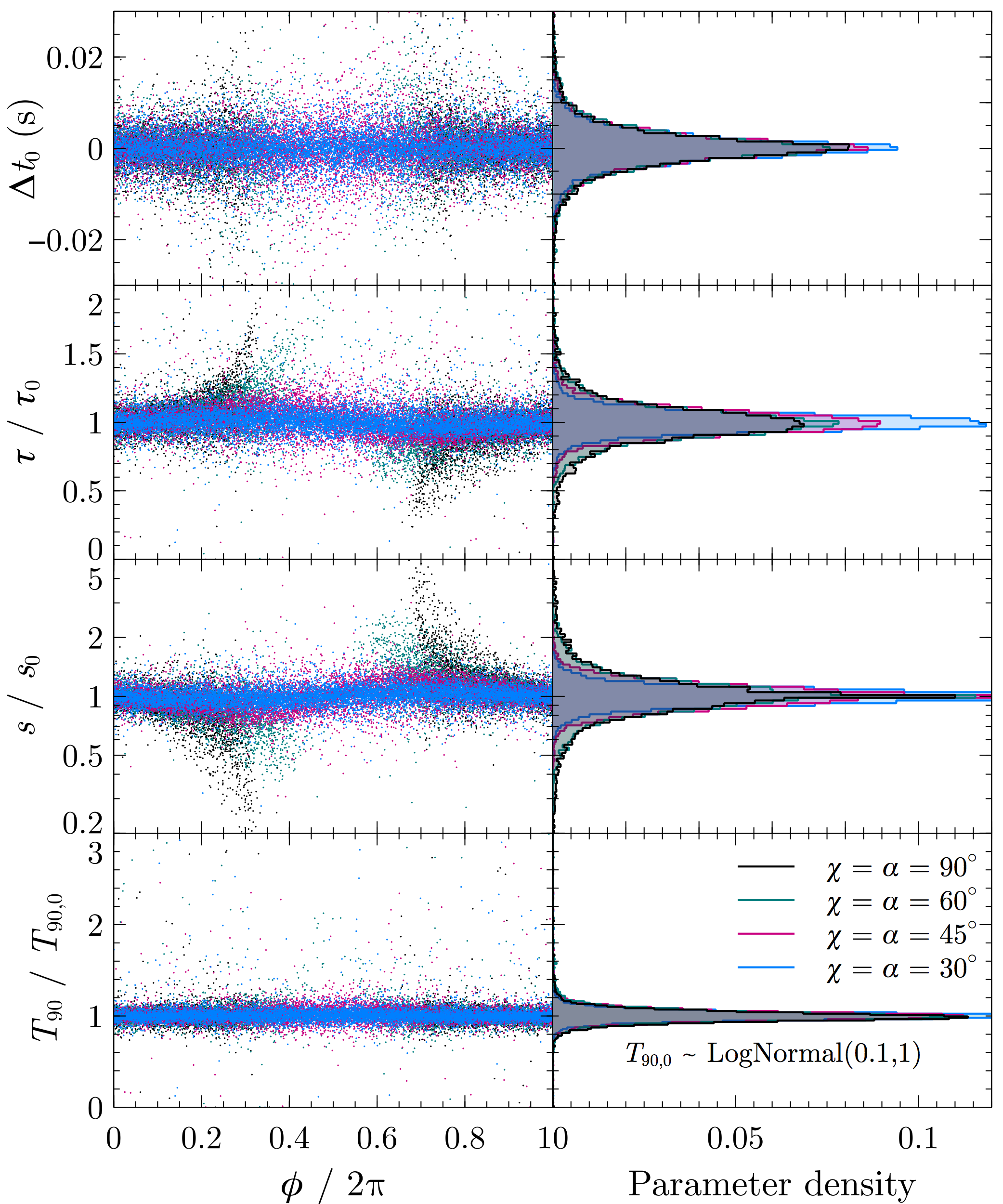}
	\caption{Phase distributions (left) and parameter densities (right) of burst parameters of Run 2, where the input burst durations are drawn from a lognormal distribution (see Fig. \ref{fig:RUN2_T90_hist}).}\label{fig:RUN2}
\end{figure}

\begin{figure}
	\centering
		\includegraphics[width=0.45 \textwidth]{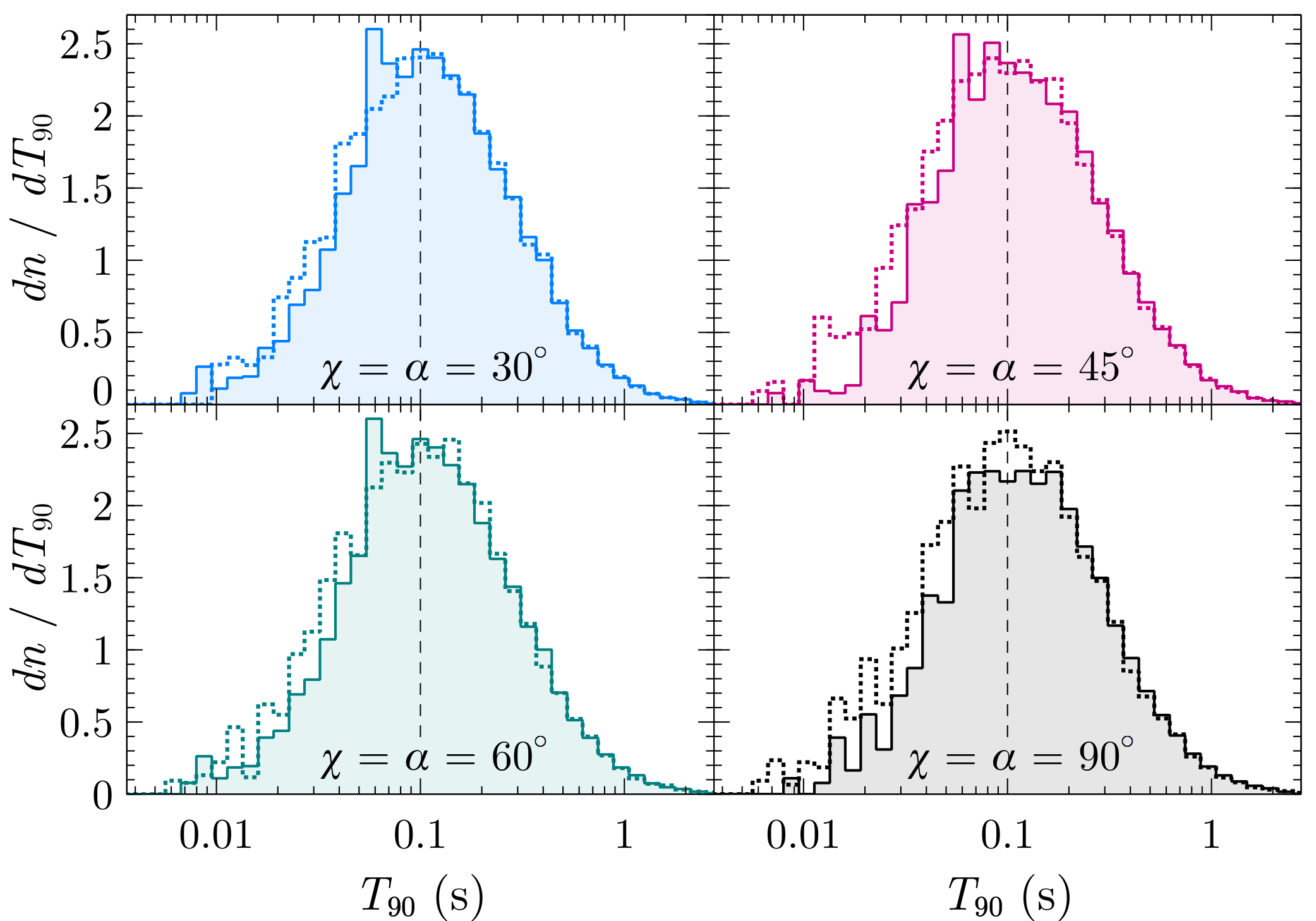}
	\caption{Burst duration $T_{90}$ distributions of 4 separate simulations (with different values for $\chi$ and $\alpha$) of Run 2. The dotted (solid) histograms represent the input (observed) burst duration distributions. The slight dearth of observed short duration bursts, present in each simulation, is simply due to the fact that they consist of fewer counts and are therefore less likely to be identified by the burst-identification algorithm.}\label{fig:RUN2_T90_hist}
\end{figure}

\begin{table}
\caption{Number of identified bursts $n_{\rm id}$ for Run 2, per configuration. The input number of bursts for each configuration was $n=10^4$.}
\centering
\begin{tabular}{lc}
\hline
$\chi=\alpha$ ($^\circ$) & $n_{\rm id}$ \\
\hline
\hline
30 & 10028 \\
45 & 9871 \\
60 & 7382 \\
90 & 5971 \\
\hline
\end{tabular}
\label{tab:sig_bursts_RUN2}
\end{table}

The results of Run 2, where we draw the burst duration for each individual burst from a lognormal distribution are shown in Fig. \ref{fig:RUN2}. Table \ref{tab:sig_bursts_RUN2} lists the number of identified bursts per configuration. The results closely resemble those of Run 1, with $T_{90}=0.15$ s (see Fig. \ref{fig:RUN1_0.15}). We only find weak phase dependencies, that only become noticeable for large values of the angles, $\chi$ and $\alpha$. In Fig. \ref{fig:RUN2_T90_hist}, the input (dotted histogram) and best-fit (solid histogram) $T_{90}$ distributions are shown for the 4 separate configurations. Notice the small dearth of short duration bursts in each histogram; short duration bursts contain fewer counts and may therefore be missed by the burst-identification algorithm (step (V) in Section \ref{sec:General simulation procedure}). We furthermore find that there is a slight excess at $T_{90}\sim0.6 s$ (although not apparent when $\chi=\alpha=90^\circ$), which is due to the fact that for most observed bursts $\tau/\tau_0<1$ and $T_{90}\propto\tau$. 

\subsubsection{Run 3: Burst amplitude distribution}\label{subsec:Run 3}

\begin{figure}
	\centering
		\includegraphics[width=0.45 \textwidth]{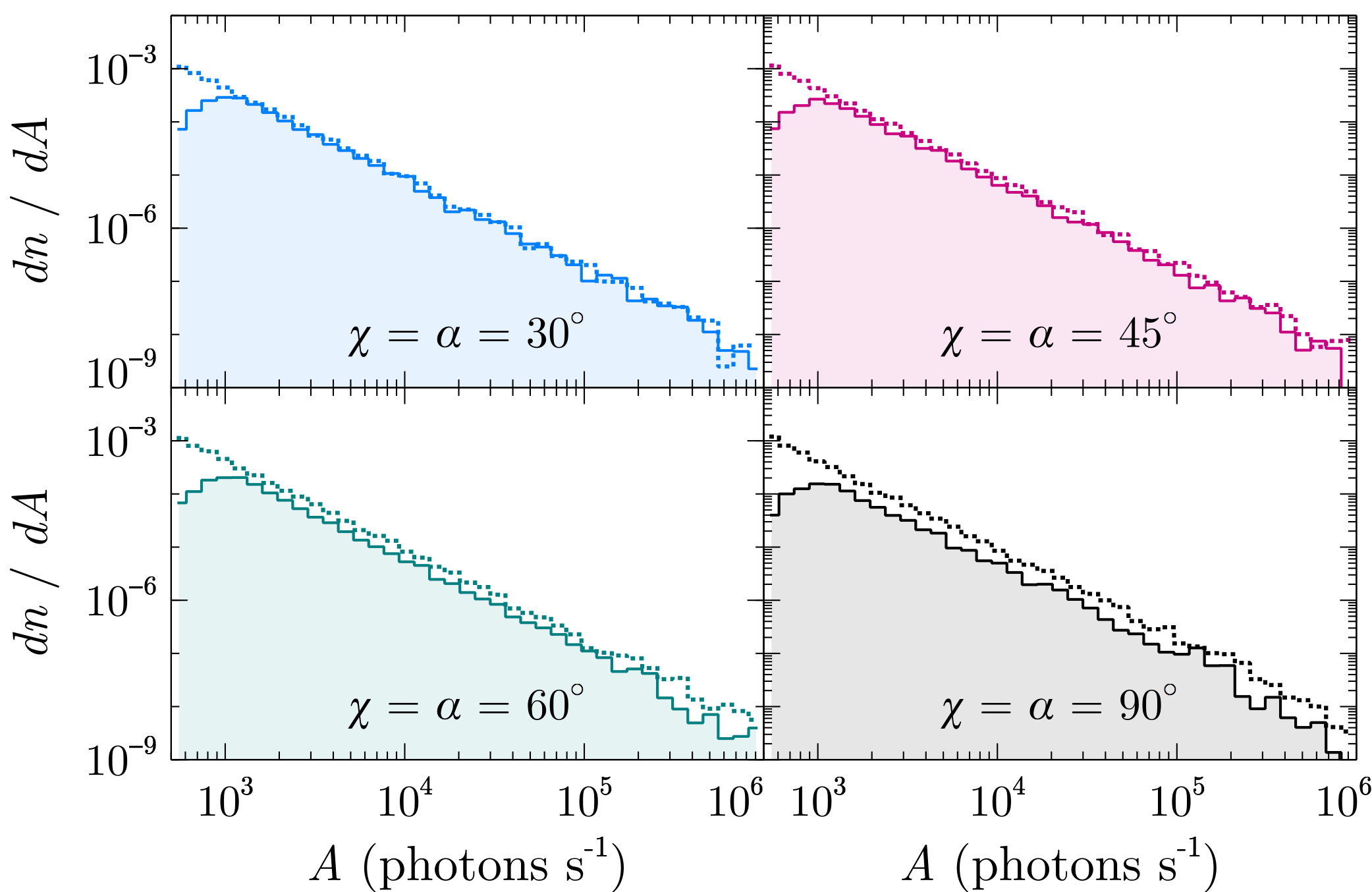}
	\caption{Burst amplitude distributions of 4 separate simulations (with different values for $\chi$ and $\alpha$) of Run 3. The dotted (solid) histograms represent the input (observed) burst amplitude distributions. The cutoff of observed low-amplitude bursts (at $\lesssim10^3$ photons s$^{-1}$) is due to the fact that the amplitude of these bursts likely occurs below the threshold of the burst-identification algorithm.}\label{fig:RUN3_y_hist}
\end{figure}

\begin{figure}
	\centering
		\includegraphics[width=0.45 \textwidth]{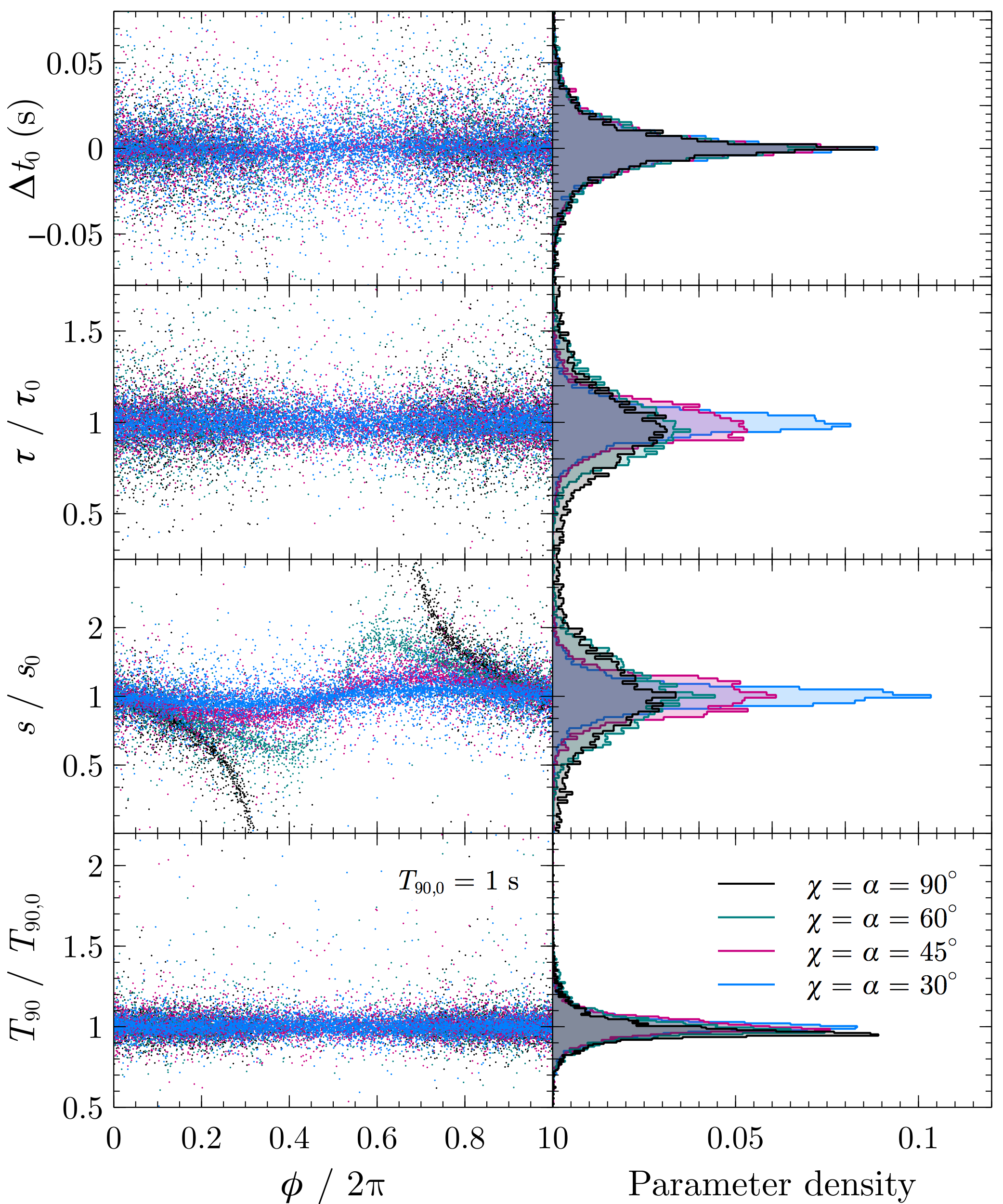}
	\caption{Phase distributions (left) and parameter densities (right) of burst parameters of Run 3, where the input burst amplitudes are drawn from a powerlaw distribution (see Fig. \ref{fig:RUN3_y_hist}).}\label{fig:RUN3}
\end{figure}

The results of Run 3, where we draw the burst amplitude/number of  emitted burst photons for each individual burst from a powerlaw distribution, are presented in Fig. \ref{fig:RUN3}. Table \ref{tab:sig_bursts_RUN3} lists the number of identified bursts per configuration. We find much less spread in the phase distributions of the parameters, compared to e.g. the results from the 1.5 s burst in Run 1 (Fig. \ref{fig:RUN1_1.50}), however we do observe a considerable amount of scatter. The latter is likely due to the fact that the majority of input bursts ($\sim0.87$) are low-amplitude bursts, i.e. $A \lesssim10^4$ photons s$^{-1}$, which are more difficult to characterize, i.e. their morphology is relatively heavily affected by Poisson noise. Fig. \ref{fig:RUN3_y_hist} displays the input and observed burst amplitude distributions. Despite a slight offset at larger angles, we find that the slope of the distributions is reproduced by the observed bursts. Input bursts with an amplitude $\lesssim10^{3}$ photons s$^{-1}$ may go unidentified as they will likely fall below the significance threshold of the burst identification algorithm. 

\begin{table}
\caption{Number of identified bursts $n_{\rm id}$ for Run 3, per configuration. The input number of bursts for each configuration was $n=10^4$.}
\centering
\begin{tabular}{lc}
\hline
$\chi=\alpha$ ($^\circ$) & $n_{\rm id}$ \\
\hline
\hline
30 & 6645 \\
45 & 5831 \\
60 & 4734 \\
90 & 3717 \\
\hline
\end{tabular}
\label{tab:sig_bursts_RUN3}
\end{table}

\subsection{Detectability of burst phase dependence}

\begin{figure}
	\centering
		\includegraphics[width=0.45 \textwidth]{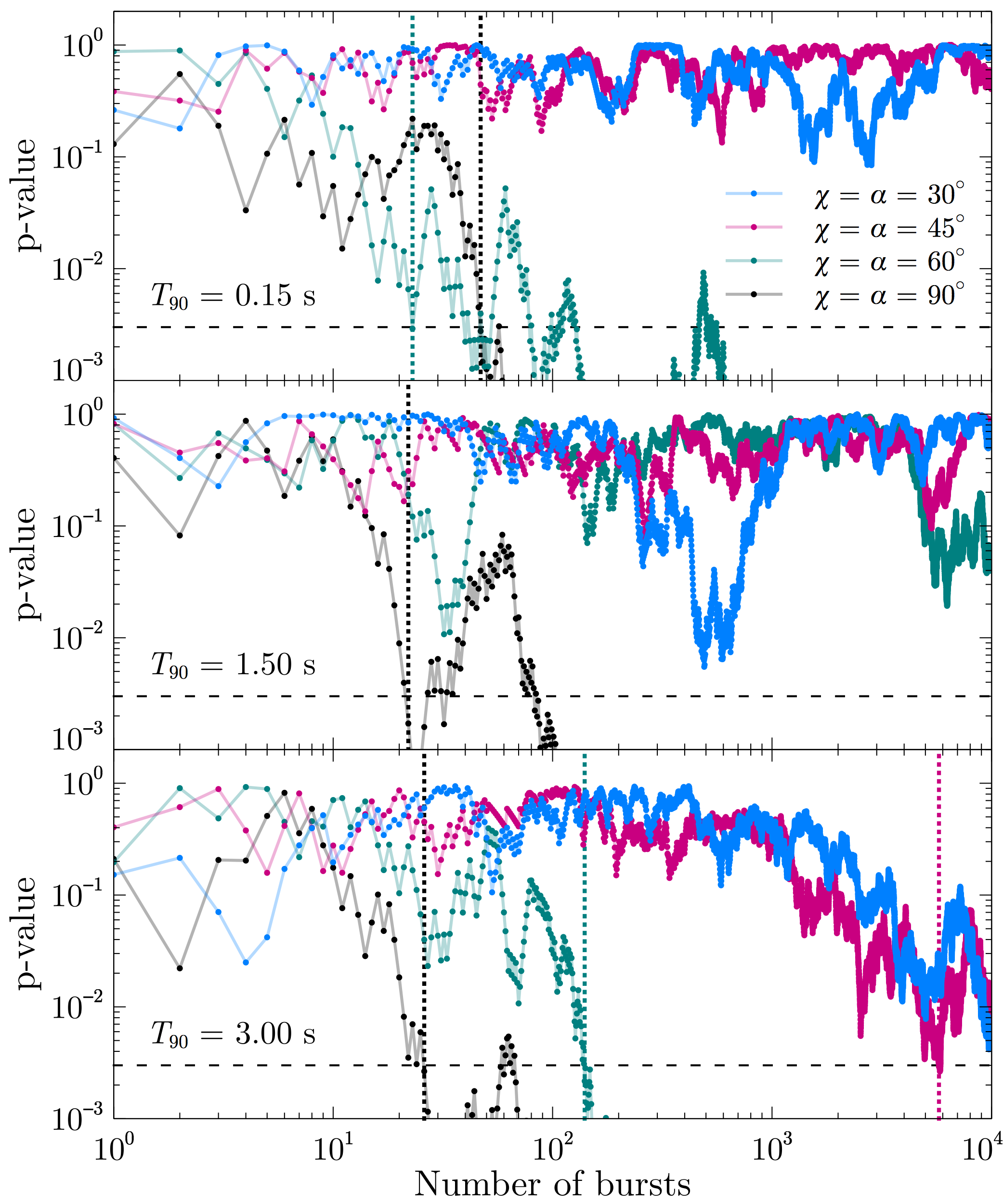}
	\caption{Evolution of the p-value against the number of bursts for 3 separate burst durations (top to bottom) of Run 1. The distinct curves per subplot represent different values for the angles $\chi$ and $\alpha$. The horizontal dashed line denotes the threshold level. The vertical dotted lines denote the number of bursts at which the p-value drops below the threshold (see text for more details).}\label{fig:RUN1_pv}
\end{figure}

Here we set out to test the main method used in studies of burst phase dependence to date, see Section \ref{sec:Overview of published burst phase-dependence analysis}. During the simulations we determine and record the phase occurrence of the burst peak $\phi_0^{\rm bf}$, i.e. the phase occurrence of the best-fit burst peak time $t_0^{\rm bf}$. After each burst we compile a distribution of the values for $\phi_0^{\rm bf}$ of all previous bursts up to the most recent one, and compare this burst phase occurrence distribution to a uniform distribution using a Kolmogorov--Smirnov (K--S) test, from which we obtain a p-value. We set the significance threshold at a p-value of 0.003, corresponding to a 0.3\% probability that the observed burst peak phase occurrences are distributed uniformly across phase. To be clear: we are simulating emission from a fixed point on the NS surface, from which bursts are being emitted at random rotational phase. The naive expectation that this would result in an observable phase-dependence most from the expected modulation of the intensity (see Fig. \ref{fig:kappa angles}) that results, for example, in missing some bursts emitted on the dark side of the star.

In Fig. \ref{fig:RUN1_pv} we plot the evolution of the p-value against the number of observed bursts up to that point in the simulation for the 3 separate burst durations (from top to bottom) of Run 1. The different curves per subplot correspond to different values of $\chi$ and $\alpha$. The horizontal dashed lines denote the threshold level and the vertical dotted lines indicate the number of bursts at which the p-value of a given simulation drops below the threshold level. Note that of these 12 simulations, the p-value does not fall below the threshold before $10^3$ bursts for $\chi=\alpha \leq 45^{\circ}$. Nevertheless, we do find a decreasing trend for $T_{90}=3.0$ s after $\sim500$ bursts, reaching the threshold at $\sim10^4$ bursts (the length of the simulation), for those angles. Remarkably, the p-value associated with the simulation where $T_{90}=1.50$ s and $\chi=\alpha=60^\circ$, does not show a decreasing trend before $10^4$ bursts. The remaining configurations do drop below the threshold fairly soon, i.e. after $\sim 20-150$ bursts.

\begin{figure*}
	\centering
		\includegraphics[width=\textwidth]{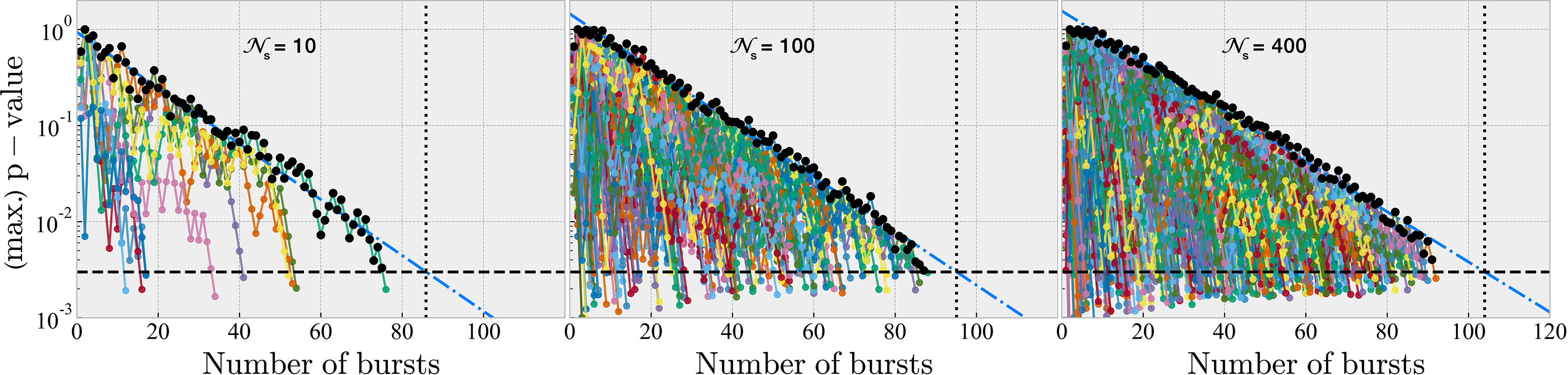}
	\caption{Evolution of the p-value for $\mathcal{N}_{\rm s}$ simulations for $T_{\rm 90}=0.15$ s and $\chi=\alpha=90^{\circ}$. From left to right we increase the value of $\mathcal{N}_{\rm s}$ from 10, to 100, to 400. The distinct coloured curves denote the evolution of the p-value for separate simulations. The black markers denote the maximum p-value that was attained (out of all simulations) for a given number of bursts. The horizontal dashed line denotes the threshold value, the cyan dash-dotted curve represents the fit to the decreasing trend of the log of the maximum p-values, and the vertical dotted curve denotes the intersection of the fit with the threshold level. The latter occurs, from left to right, at $n_{\rm min}$ = 86, 95, and 104, respectively. The maximum p-values in the panel on the right are also plotted in the top panel of Fig. \ref{fig:pv_limits}.}\label{fig:(max)_pv_evolution}
\end{figure*}

\begin{figure}
	\centering
		\includegraphics[width=0.45 \textwidth]{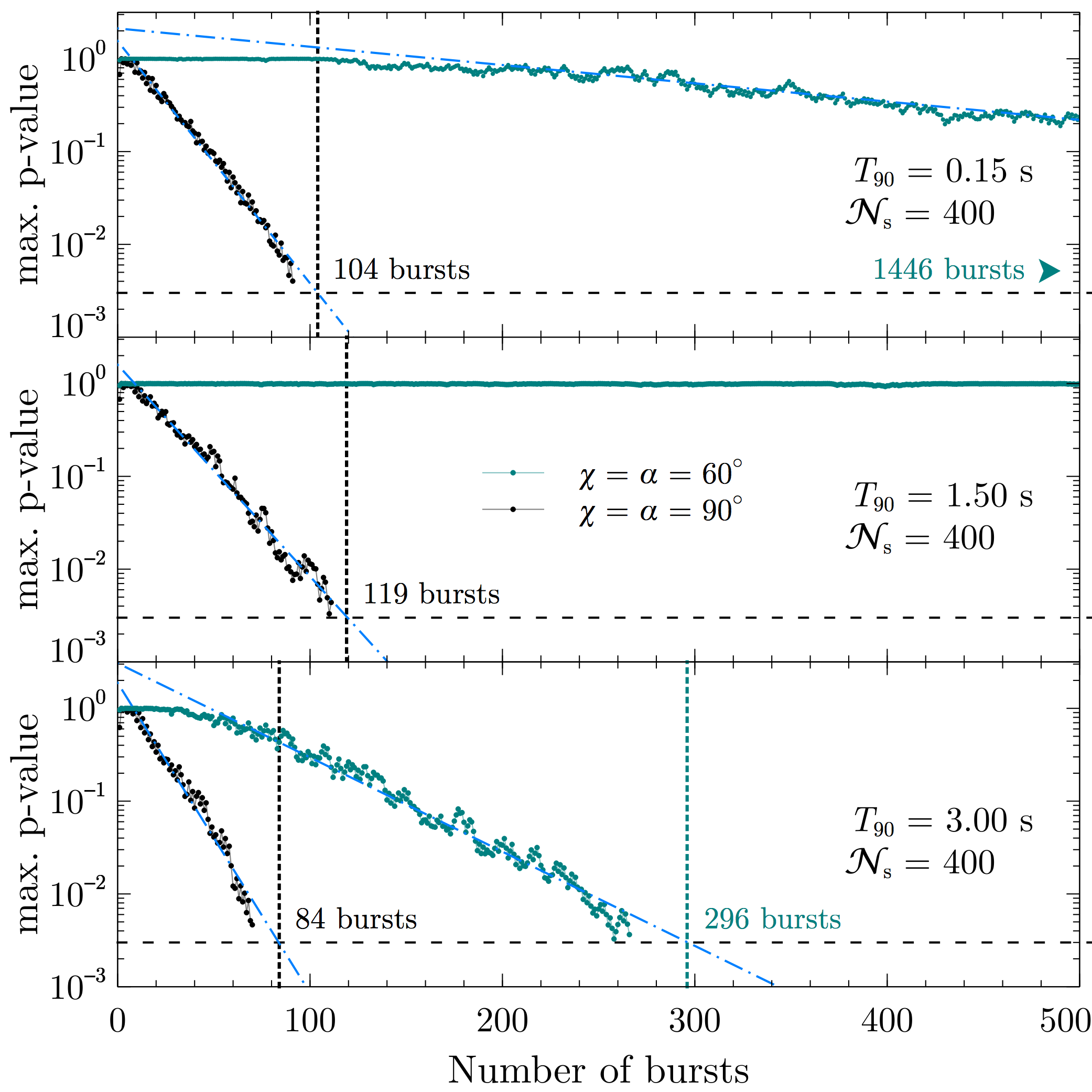}
	\caption{Maximum p-value attained, out of $\mathcal{N}_{\rm s}$ simulations per configuration, against the number of bursts for the 3 separate burst durations (top to bottom) of Run 1. The horizontal dashed line denotes the threshold level, the blue dash-dotted curves represent the fits to the decreasing trends, and the vertical dotted lines denote the number of bursts where the fits intersect with the threshold level. }\label{fig:pv_limits}
\end{figure}

To determine the minimum number of bursts required to guarantee that the p-value drops below the threshold, we ran the simulation per configuration $\mathcal{N}_{\rm s}$ times, each time until the threshold was reached, and recorded the number of bursts. Fig. \ref{fig:(max)_pv_evolution} shows the evolution of the p-value for $\mathcal{N}_{\rm s}$ = 10, 100, and 400 simulations, with $T_{\rm 90}=0.15$ s and $\chi=\alpha=90^{\circ}$. We found that the maximum obtained p-value (out of all $\mathcal{N}_{\rm s}$ simulations for a given configuration) after a specific number of bursts decreases at a certain rate (denoted by the black markers). In  Fig. \ref{fig:pv_limits}, we plot the maximum p-value attained, over $\mathcal{N}_{\rm s}=400$ simulations per configuration, against the number of bursts. Subsequently, we fit a straight lines to the decreasing trends of the log of the p-value and record the number of bursts at which these lines intersect with the threshold level. Accordingly, we find an estimate for the minimum number of bursts $n_{\rm min}$ at which, assuming a certain configuration, the observed $\phi_0^{\rm bf}$ distribution should deviate significantly from a uniform distribution. If the $\phi_0^{\rm bf}$ distribution does not significantly deviate from a uniform distribution after $n_{\rm min}$ bursts, then the configuration will likely be such that the modulation in intensity is less strong than assumed. The latter is dependent on assumptions on the parameters that determine the shape of $\kappa(\theta_0)$ (equation \ref{eq:kappa}). 

For the burst durations and configurations that we study in Run 1, we find for $\chi=\alpha = 90^\circ$ that $n_{\rm min}\sim100$ bursts. For $\chi=\alpha = 60^\circ$, we find $n_{\rm min}=1446$ bursts and $n_{\rm min}=296$ bursts, for $T_{90}=0.15$ s and $T_{90}=3.0$ s, respectively. Yet, we do not find an $n_{\rm min}$ for $T_{90}=1.50$ s, since the attained maximum p-value does not exhibit a decreasing trend before $10^3$ bursts; consistent with the simulation run displayed in the middle panel of Fig. \ref{fig:RUN1_pv}. This is because the burst spot remains (partially) visible throughout the NS's rotation, such that enough counts can be detected for the duration of the burst, and the fact that the $\Delta t_{0}$ remains comparatively small, i.e. the corresponding parameter density comprises a narrow peak (Fig. \ref{fig:RUN1_1.50}), in contrast to e.g. the parameter density of the 3.0 s burst, which is much more spread out (Fig. \ref{fig:RUN1_3.00}).

\section{Discussion and Conclusion}\label{sec:Discussion and Conclusion}

\begin{figure}
	\centering
		\includegraphics[width=0.4 \textwidth]{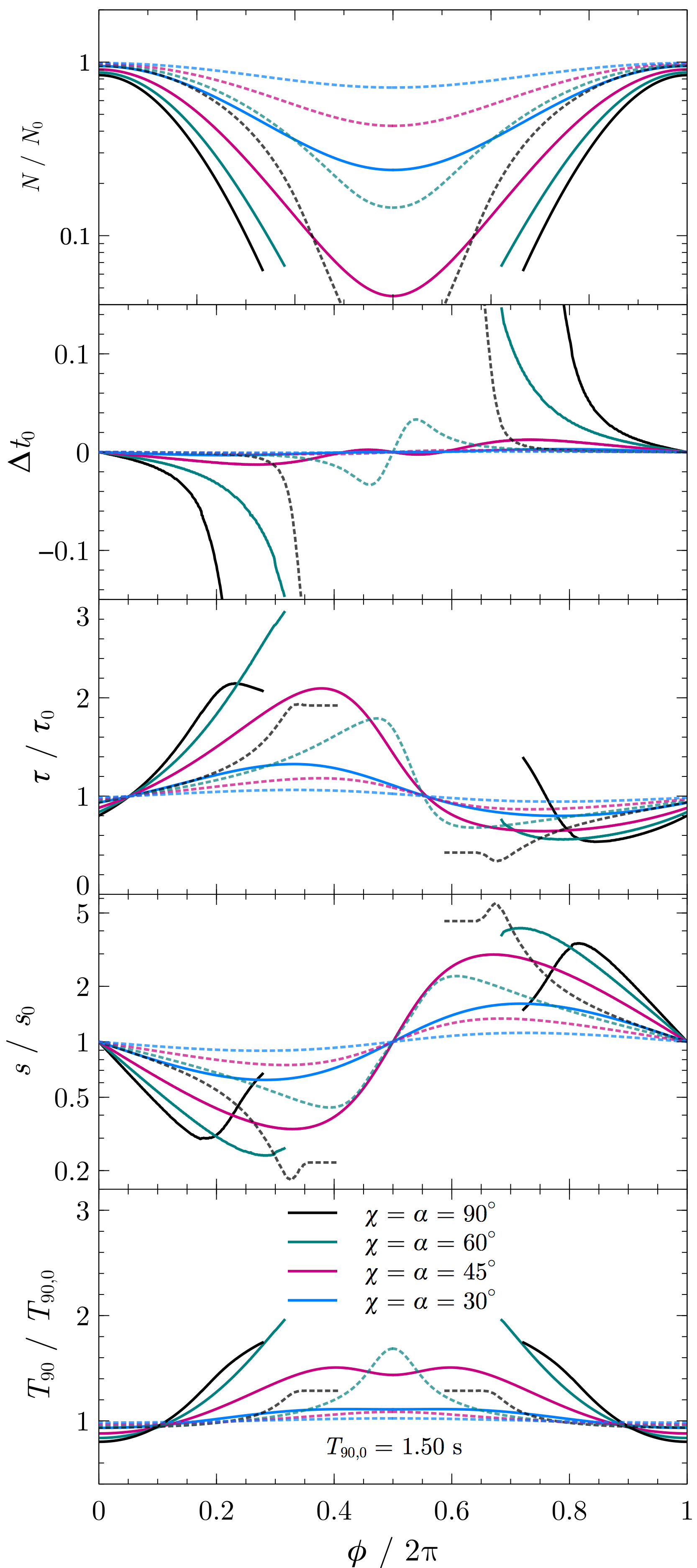}
	\caption{Predicted phase distributions of burst parameters in the presence of beamed emission for a burst of $T_{90}=1.5$ s (solid curves). Beaming is described by equation (\ref{eq:beaming function}), where we have set the beaming width $\sigma_{\rm b}$ to $\pi/6$. For comparison, the dotted curves represent the burst parameter distributions for an isotropic burst with the same duration.}\label{fig:RUN1_analytic_beam}
\end{figure}

We have studied, from a theoretical perspective, the conditions under which magnetar bursts from a predefined localized active region or burst patch on the NS surface would give rise to a detectable phase-dependence. By adopting a straightforward input burst model, we were able to examine the changes in the observed bursts after they were modulated by the phase-dependent function $\kappa_*(\phi)$, which takes into account the effects of gravitational light bending and depends on the configuration of the system. 

We found that the degree to which the inferred burst properties become phase dependent is strongly contingent on the duration of the bursts and geometry of the system; we find a stronger phase dependency of the burst properties for longer duration bursts and larger values of the angles $\chi$ and $\alpha$. The former is because longer bursts sample a wider range of photon trajectories and the latter and the latter is due to the fact that for larger values of $\chi$ and $\alpha$ the GR effects become more significant. Furthermore, the majority of observed bursts turn out to have $\tau/\tau_0<1$ and $T_{90}/T_{90,0}<1$, i.e. they rise faster and appear shorter than their input counterparts. Attempts to infer the properties of individual bursts with durations greater than $\sim 10-20$ \% of the spin period should certainly take into account potential distortion due to phase-dependent effects.

Adopting a lognormal burst duration distribution that peaks at $\overline{T}_{90}=0.1$ s (as observed for well-sampled sources), from which we draw the input duration for each individual burst, we found that phase distributions of the parameters closely resembled those of Run 1, for which $T_{90}=0.15$ s. When considering a powerlaw distribution for the input burst amplitudes and burst duration of $T_{90}=1$ s, we observed a weak phase-dependency of the burst parameters and a considerable amount of scatter, which in turn is caused by the large fraction of low-amplitude input bursts, which are more affected by Poisson noise. We conclude that the observed distributions of burst properties from well-sampled sources are likely not strongly distorted due to phase-dependent effects, by virtue of being dominated by short bursts.

We studied the detectability of phase-dependence, using the most commonly-used measure (see Section \ref{sec:Overview of published burst phase-dependence analysis}) whereby one concentrates on the phase occurrence of the burst peaks. In our setup all bursts originate at a specific small active region -- in some respects the most extreme phase-dependent scenario. However rotational phase dependence of the peak occurrences was not always apparent. We found that one would require a minimum number of bursts for certain input burst properties and a given system configuration, to \emph{guarantee} observing a phase dependence. Only in the case of the most extreme geometries, i.e. $\chi=\alpha>60^\circ$, does this approach the burst sample sizes that were examined in the literature, which range from tens to several hundred bursts. Studies that have not found a phase-dependence in the distribution of the burst peak occurrences as yet \citep[e.g.][]{Savchenko2010,Scholz2011,Collazzi2015}, might simply require a larger burst sample in order rule it out for certain geometries.  For other geometries, however, it will never be possible to rule out the presence of a burst phase-dependence.

In our study we have considered only a restricted range of scenarios, where the emission region is tied to the stellar surface. One factor that we have not simulated in detail is that of any potential beaming of the burst emission. To offer brief insights for such influences, we show in Fig. \ref{fig:RUN1_analytic_beam} the theoretical phase distributions of the observed burst properties in the case of beamed emission (Equation \ref{eq:beaming function} with $\sigma_{\rm b}=\pi/6$) from a burst with $T_{90}=1.5$ s; the corresponding shape of $\kappa_*(\phi)$ is shown by the dashed curves in Fig. \ref{fig:kappa angles}. We compare it to its isotropic counterpart and find that the phase-dependency of the burst properties will be enhanced in the presence of beaming. 

The detectability of a burst phase-dependence depends strongly on the shape of $\kappa_*(\phi)$: the stronger the variation with $\phi$ the greater the modification to the input burst profiles. Introducing additional bursts patches or allowing for active regions to occur at a certain height above the surface, will cause the phase-dependence of $\kappa_*$ to decrease. A burst phase-dependence in those cases may then only become detectable if the emission is also strongly beamed. 

In this paper we have not studied the method whereby a phase-dependence is searched for in the epoch folded photon times of arrival.  This method can, and should, be subjected to the same level of scrutiny. An additional challenge with this method, however, is to determine a proper false alarm rate. Straightforwardly looking for deviations from uniformity of the times of arrival does not work, since a single burst already consists a significant departure. One must instead quantify the conditions under which one would detect a burst photon phase-dependence, even if the bursts originated at random locations on or above the NS surface. We defer this topic to future studies.

\section*{Acknowledgements}
CE and ALW acknowledge support from NOVA, the Dutch Top Research School for Astronomy. DH was partially supported by the Moore-Sloan Data Science Environment at New York University, and the James Arthur Postdoctoral Fellowship at New York University. DH acknowledges support from the DIRAC Institute in the Department of Astronomy at the University of Washington. The DIRAC Institute is supported through generous gifts from the Charles and Lisa Simonyi Fund for Arts and Sciences, and the Washington Research Foundation. We thank Matthew Baring for useful comments. 

\bibliographystyle{mnras}
\bibliography{biblio}

\appendix

\section{Parameter table}

Here we list brief descriptions of the simulation parameters and their associated symbols (see Table \ref{tab:sim_pars}).  

\begin{table}
\caption{A table of the simulation parameters appearing in sections \ref{sec:Simulations} and \ref{sec:Results}.}
\centering
\begin{tabular}{ll}
\hline
Symbol & Description \\
\hline
\hline
$n$ & number of input bursts\\
$n_{\rm id}$ & number of identified bursts\\
$n_{\rm min}$ & number of bursts at which the p-value is\\
 & guaranteed to drop below the threshold \\
$R$ & neutron star radius\\
$P$ & neutron star rotation period\\
$\psi$ & size of the burst patch\\
$\chi$ & angle between NS axis of rotation and \\
& the line-of-sight \\
$\alpha$ & colatitude of the burst patch\\
$\zeta$ & background rate\\
$\delta t$ & light curve bin width\\
$b$ & background level\\
$N$ & number of emitted photons\\
$N_{\rm det}$ & number of detected photons\\
$t_0$ & burst peak time\\
$\phi_0$ & burst peak phase occurrence\\
$\tau$ & rise-time\\
$s$ & skewness factor\\
$A$ & burst amplitude\\
$\mathcal{N}_{\rm bins}$ & number of time bins\\
$y^{\rm sig}$ & maximum departure amplitude\\
$t_0^{\rm sig}$ & time bin with $y^{\rm sig}$\\
$\Delta T$ & time interval of an observed burst\\
$t_{\rm in}$, $t_{\rm out}$ & limits of $\Delta T$\\
$\mu$ & mean count level $\sim b$\\
$\mu^*$ & running mean\\
$\Delta T_{\rm interval}$ & time interval over which $\mu^*$ is estimated\\
$\Delta t_0$ & difference input and best-fit burst peak\\ 
& time: $\Delta t_0\equiv t^{\rm bf}-t_0$\\
$T_{90}$ & burst duration\\
$\overline T_{90}$ & mode of the duration distribution\\
$\sigma_{T_{90}}$ & width of the duration distribution\\
$T_{90}^{\rm min}$ & minimum burst duration\\
$T_{90}^{\rm max}$ & maximum burst duration\\
$\Gamma$ & powerlaw index of the amplitude\\
&distribution\\
$A^{\rm min}$ & minimum burst amplitude\\
$A^{\rm max}$ & maximum burst amplitude\\
$\mathcal{N}_{\rm s}$ & number of simulations\\
\hline
subscript `0' & input parameter \\
superscript `init' & initial guess \\
superscript `bf' & best-fit parameter \\
\hline
\end{tabular}
\label{tab:sim_pars}
\end{table}

\bsp	
\label{lastpage}
\end{document}